%% file: CEoptim_arxiv.tex
\documentclass[article,shortnames,nojss]{jss}

\usepackage{float}
\usepackage{amsmath}
\usepackage{amsfonts}
\usepackage{amssymb}
\usepackage{mathrsfs}
\usepackage{euscript}
\usepackage{longtable}  %break page
\usepackage{algorithm}
\usepackage{algorithmic}
\usepackage{graphicx}

\include{definitions}

\author{Tim Benham\\The University of \\Queensland, Brisbane \And
Qibin Duan\\The University of \\Queensland, Brisbane \And
Dirk P. Kroese\\The University of \\Queensland, Brisbane \And
Beno\^{i}t Liquet\\The University of \\Queensland, Brisbane}

\Plainauthor{Tim Benham,
 Qibin Duan,
 Dirk P. Kroese,
  Beno\^{i}t Liquet,
 }

\title{\pkg{CEoptim}: Cross-Entropy \proglang{R} Package for Optimization}

\Plaintitle{CEoptim: Cross-Entropy R Package for Optimization}

\Shorttitle{\pkg{CEoptim}: Cross-Entropy \proglang{R} package for Optimization}

\Abstract{The cross-entropy (CE) method is simple and versatile
  technique for
  optimization, based on Kullback-Leibler (or cross-entropy)
 minimization.  The method can be applied to a wide
 range of optimization tasks, including continuous, discrete, mixed
 and constrained optimization problems. The new package
  \pkg{CEoptim} provides the \proglang{R} implementation of the CE
method for optimization. We describe the general CE methodology for
optimization and well as some useful
  modifications. The usage and efficacy of   \pkg{CEoptim} is
  demonstrated through a variety of
optimization examples,
  including model fitting, combinatorial optimization, and maximum likelihood
  estimation.  }

\Plainkeywords{Cross Entropy, Optimization, R}

\Address{
School of Mathemathics and Physics\\
The University of Queensland\\
Brisbane, Australia\\
E-mails: \\
\email{tim.j.benham@gmail.com}\\
\email{q.duan@uq.edu.au}\\
\email{kroese@maths.uq.edu.au} (corresponding author)\\
\email{b.liquet@uq.edu.au}
}

\usepackage{verbatim}
\usepackage{dsfont}

\begin{document}

\section{Introduction}
The cross-entropy (CE) method originates from an adaptive variance
minimization algorithm in
\cite{rubinstein1997} for the estimation  rare event probabilities in stochastic networks. It was realized  in
\cite{rubinstein1999} that many
optimization
problems could be converted into a rare-event estimation problems,
providing a rare-event based approach to optimization, where a
sequence of
probability densities is generated that converges to a degenerate
density
that concentrates its mass close to the optimizer.

Generally, the CE method involves two iterative phases:
\begin{enumerate}
\item Generation of a set of random samples (vectors, trajectories,
  etc.) according to a specified parameterized model.
\item Updating of the model parameters, based on the best
  samples generated in the previous step. This is done by
  Kullback--Leibler (also called cross-entropy) minimization.
\end{enumerate}

%CE  method has been implemented as a \proglang{MATLAB} toolbox
%\proglang{CEToolBox},which is available at
%\url{http://www.maths.uq.edu.au/CEToolBox/}.
Since the appearance of the CE monograph \citep{cebook} and the
tutorial \citep{cetutorial2005}, the CE method has continued
to develop and has been successfully applied to a great variety of
difficult optimization problems, including
%{\color{red} reference about CE optimization will be listed here}
 motion
planning in robotic systems \citep{kobilarov2012motionplanning},
electricity network generation,
 \citep{kothari09integerprograming}, control of infectious
 diseases
 \citep{sani2008controlling}, buffer allocation
\citep{alon2005allocation}, Laguerre tessellation
\citep{duan2014inverting}, and network
reliability \citep{kroese2007network}. An extensive list of recent
work can be found in \citep{ceoptbotev2013}. Websites that provide
$\matlab$ code include \url{www.cemethod.org} and
\url{www.montecarlohandbook.org}.
Since \proglang{R} has become an essential tool for
statistical computation, it is useful to provide an accessible
implementation of the CE method for \proglang{R} users, similar to
 \proglang{R} packages for
simulated annealing \citep{xiang2013sa}, evolutionary methods
\citep{mullendeoptimJss}, and particle swarm optimization
methods \citep{bendtsenpso}.

Some advantages of the CE method are:
\begin{itemize}
\item The CE method is a global optimization method which is
  particularly useful when the objective function has many local
  optima.
\item The CE method can be used to solve continuous, discrete, and mixed optimization problems, which may also include
  constraints.
\item The CE code is extremely compact and is readily written in native
\proglang{R}, making  further development and modifications easy to
implement.
\item The CE method is based on rigorous mathematical and statistical
 principles.
\end{itemize}

Our aim is not to replace the standard optimization solvers such
as \pkg{optim} and \pkg{nlm}
but to provide a viable alternative in cases where standard gradient or
simplex-based solvers are not applicable (e.g., when the optimization
problem contains both discrete and continuous variables) or are expected to
do poorly (e.g., when there are many local optima).

%%  Our motivation here is to extend the set of
%% algorithms for global optimization problems in \proglang{R}
%% platform. One obvious advantage of \proglang{R} is open-sourced, which
%% means users and developers of \proglang{R} language and environment do
%% not have to pay for the software, making more people have access to
%% \proglang{R}. A implementation of CE method in \proglang{R} can save
%% much time and effort for people who want to use CE method in
%% \proglang{R} environment. Moreover, \proglang{R} has an increasingly
%% amount of users interested in optimization research, which requires
%% more kinds of optimization packages.

The rest of this paper is organized as follows. In Section~\ref{sec:cealg}, we sketch the general theory behind the CE method,
which leads to the basic CE algorithm. In Section~\ref{sec:ceropt}, we
describe a variety of optimization scenarios,
including continuous, discrete and constrained mixed problems, to
which CE can be applied effectively. The
description and usage of the \pkg{CEoptim} package
are given in Section \ref{sec:ceusage}. Section~\ref{sec:numerical}
demonstrates the capability of the package through a range of numerical examples. In the final section we make concluding remarks for
\pkg{CEoptim}.
%.....{\color{red} other sections will be described}

\section{CE method for optimization} \label{sec:cealg}

Let $\scX$ be an arbitrary set of {\em states} and let $S$ be a
real-valued performance function on $\scX$.  Suppose the goal is to find
the minimum of $S$ over $\scX$, and the corresponding minimizer
$\bx^*$ (assuming, for simplicity, that there is
only one). Denote the minimum by $\gamma^*$, so that
\begin{equation}\label{prob1}
S(\bx^*) = \gamma^* = \min_{\bx \in \scX} S(\bx).
\end{equation}

% This setting includes many types of optimization problems: discrete
% (combinatorial), continuous, mixed, and constrained problems.
% Moreover, if one is interested in
% minimizing rather than maximizing $S$, one can simply maximize $-S$.

The CE  methodology for optimization is adapted from the CE
methodology  for rare
event estimation in the following way.
Associate with the above problem \eqref{prob1} the estimation of the probability
$\ell = \Pm(S(\bX) \leq \gamma)$, where $\bX$ has some probability
density $f(\bx;\bu)$ on $\scX$ (for example corresponding to the
uniform distribution on $\scX$) depending on a parameter $\bu$ and a
{\em level} $\gamma$.
Thus, for optimization problems randomness is purposely introduced in order to make the model
stochastic.
If
$\gamma$ is chosen close to the unknown $\gamma^*$, then   $\ell$
is typically a rare-event
probability. One of the most effective ways to estimate rare-event
probabilities is to use {\em importance sampling}.
In particular, to estimate $\ell = \Pm(S(\bX) \leq
\gamma)$ one can use the importance sampling estimator
\[
\hat{\ell} = \frac{1}{N}\sum_{i=1}^N \frac{f(\bX_i)}{g(\bX_i)} \I\{S(\bX_i)
\leq \gamma\},
\]
where $\bX_1,\ldots,\bX_N$ are iid samples from a well-chosen
importance sampling density $g$. The optimal importance sampling
density is in this case $g^*(\bx) =  f(\bx) \I\{S(\bx) \leq
\gamma\}/\ell$, which gives a zero-variance estimator, but depends on
the unknown quantity $\ell$. The main idea behind the CE method for
estimation is to adaptively determine an importance sampling pdf
$f(\bx; \bv^*)$ --- hence within the same family as the original
distribution --- that is close to $g^*$ in Kullback--Leibler
sense. Specifically, a parameter $\bv^*$ is sought that minimizes
the cross-entropy distance
\[
\KL(g^*, f(\cdot;\bv)) = \Em_{g^*}\left[ \log \frac{g^*(\bX)}{f(\bX; \bv)}\right] =
\int g^*(\bx) \log g^*(\bx) \,\di \bx - \int g^*(\bx) \log f(\bx;\bv) \, \di
\bx\;.
\]
This is equivalent to maximizing, with respect to $\bv$,
\[
\int f(\bx;\bu)\I\{S(\bx) \leq \gamma\} \log f(\bx; \bv)\, \di \bx =
\Em_{\bu}\left[  \I\{S(\bX) \leq \gamma\} \log f(\bX; \bv)\right]\;,
\]
which in turn can be estimated by maximizing the sample average
\begin{equation}\label{sampav}
\frac{1}{N} \sum_{i=1}^N \left[  \I\{S(\bX_i) \leq \gamma\} \log f(\bX_i; \bv)\right]\;,
\end{equation}
where $\bX_1, \ldots,\bX_N$ is an iid sample from $f(\bx;\bu)$.
This is, in essence, maximum likelihood estimation. In
particular, \eqref{sampav} gives the maximum likelihood estimator of
$\bv$ based on only the samples $\bX_1,\ldots,\bX_N$ that have a
function value less than or equal to $\gamma$. These are the so-called
{\em elite samples}.

The relevance to optimization is that when $\gamma$ is close to the
(usually unknown) minimum $\gamma^*$, then the importance sampling
density $g^*$ concentrates
most of its  mass in the vicinity of the minimizer  $\bx^*$.
Sampling from such a distribution
thus produces optimal or near-optimal states.
 The CE method for optimization   produces
a sequence of levels $({\gamma}_t)$ and reference parameters
$({\bv}_t)$ determined from \eqref{sampav}
%(where, for each
%iteration $t$,  $\gamma = \gamma_t$
%and $\bu = \bv_{t-1}$)
such that the former tends to the optimal $\gamma^*$
and the latter to the optimal reference vector $\bv^*$, where
$f(\bx; \bv^*)$ corresponds to
the point mass at $\bx^*$; see, e.g., \cite[Page 251]{mcbook}.

 The  generic steps for CE optimization are specified in Algorithm~\ref{alg:CE}.

\begin{algorithm}[H]
\medskip

\caption{Generic CE algorithm}
\begin{algorithmic}[1]
\REQUIRE  Initial parameter vector  $\bv_0$. Sample size
$N$. Rarity parameter $\rho$.
\ENSURE Sequence of levels $({\gamma}_t)_{t=1}^T$ and parameters
  $({\bv}_t)_{t=1}^T$.
\STATE
Let $N^{\elite} = \ceil{\rho N}$ (number of elite samples) and set $ t = 1$
(level counter).
\WHILE {the sampling distribution is not degenerate}
\STATE  Generate $\bX_1,
\ldots, \bX_N \simiidt f(\cdot; {\bv}_{t-1})$.
Calculate
 the performances $S(\bX_i)$ for all $i$, and order them from
 smallest to largest: $S_{(1)}\leq \ldots \leq S_{(N)}$. Let
 ${\gamma}_t$ be the sample $\rho$-quantile of
 performances; that is,  ${\gamma}_t = S_{(N^{\elite})}$.
\STATE Use the {\bf same}  sample
$\bX_1,\ldots,\bX_N$
  and solve  the
 stochastic program \begin{equation}\label{eq:stcp2}
\max_{\bv} \sum_{k=1}^N
 \I\{S(\bX_k) \leq {\gamma}_t\} \,  \log f(\bX_k;\bv)\;.
\end{equation}
Denote the solution   by ${\bv}_t$. Increase $t$ by 1.
\ENDWHILE
\end{algorithmic}
\label{alg:CE}
\end{algorithm}

\medskip

To run the algorithm, one needs to provide the class of sampling
densities $\{f(\cdot;\bv)\}$,
the initial vector ${\bv}_0$, the sample size $N$, the rarity
parameter $\rho$, and the stopping criterion.
It is prudent to keep track of the overall
best function value and corresponding state, and report these at the
end of the algorithm as the optimal value
and optimizer, respectively. The progression of level parameter
${\gamma}_t$ gives an indication how well the algorithm
converges.

As  \eqref{eq:stcp2}
is simply a maximum likelihood
estimation step involving only the elite samples, it is possible to derive
easy parameter updates for standard sampling distributions.
The following two special cases are of particular importance.
\begin{enumerate}
\item {\bf Multivariate normal distribution.} Suppose each $\bX$ is sampled from
  an $n$-dimensional multivariate normal distribution with independent
  components. The parameter vector $\bv$ in the CE algorithm can be
  taken as the $2n$-dimensional vector of means and standard deviations. In each
  iteration these means and standard deviations are updated according
  to the sample mean and sample standard deviation of the elite
  samples.

\item {\bf Multivariate Bernoulli distribution.}  Suppose each $\bX$ is sampled from
  an $n${-\penalty0\hskip0pt\relax}dimensional Bernoulli distribution with independent
  components. The parameter vector $\bv$ in the CE algorithm can be
  taken as the $n$-dimensional vector of success probabilities. In each
  iteration the $i$th success probability is updated according to the
  mean number of successes (1s) at the $i$th position of the elite
  samples.
\end{enumerate}

\begin{remark}[Parameter Smoothing]\rm
Various modifications of the basic CE algorithm have been proposed in
recent years.
One such is modification is {\em parameter smoothing}, where at the
$t$th iteration the sampling
parameter is updated  via
\begin{equation}\label{smoothupd2}
{\bv}_{t} = \alpha \, \widetilde{\bv}_{t} + (1- \alpha)\,  {\bv}_{t-1},
\end{equation}
where $\widetilde{\bv}_{t}$ is the solution to  \eqref{eq:stcp2} and
$0\leq \alpha\leq 1$ is a fixed smoothing parameter.

Smoothed updating can prevent the sampling distribution from
converging too quickly to a sub-optimal
degenerate distribution. This is especially relevant for the
multivariate Bernoulli case where, once a success probability reaches
0 or 1, it can no longer change.

It is also possible to use different smoothing parameters for
different components of the parameter vector (e.g., the means and
the variances).
%% For example, in \cite{kroese2006} the following
%% dynamic smoothing scheme is introduced:
%% \begin{equation}\label{dynamicsmooth}
%% \beta_t=\beta-\beta\left(1-\dfrac{1}{t}\right)^q,
%% \end{equation}
%% where $\beta$ is the initial smoothing parameter, and $q$ controls the converge speed of $\beta_t$.

%% \intrusion{Are we going to implement this or ``inflation''?}
%% Another way to prevent premature shrinkage of the sampling
%% distribution is to use {\em injection}
%% \citep{botev2004mixture}. The idea, for multivariate normal sampling,
%% is as follows.  Let $S^*_t$  denote the best performance found at
%% iteration $t$, and let  $\sigma^*_t$ be the largest standard deviation
%% at $t$.  If  $\sigma^*_t$ is sufficiently small and $\vert
%% S^*_t-S^*_{t-1} \vert$ is also very small, then add $B=c\vert
%% S^*_t-S^*_{t-1} \vert$ to each standard deviation, for some fixed
%% $c$. Alternatively, just some constant value is added to the standard variance.
\end{remark}
\begin{remark}[Choice of sampling densities]\rm
Although sampling distributions with independent components are the
most convenient to use in a CE implementation, it is sometimes
advantageous consider more complex sampling models, such as mixture
models. In this case the updating of parameters (maximum likelihood
estimation) may no longer be
trivial, but one can instead employ fast methods such as the EM algorithm
to determine the parameter updates.
\end{remark}

\begin{remark}[Choice of the CE parameters]\rm
The CE method is fairly robust with respect to the choice of the
parameters. The rarity parameter $\rho$ is typically chosen between 0.01 and
0.1. The number of elite samples $N^{\elite} = \ceil{\rho N}$ should be large
enough to obtain a reliable parameter update in \eqref{eq:stcp2}. For
example, if the dimension of $\bv$ is $d$, the number of elites should
be in the order of $10 \,d$ or higher.
\end{remark}

\section{Optimization scenarios} \label{sec:ceropt}
In this section we consider a number  optimization scenarios
to which
\pkg{CEoptim} could  be applied.

\subsection{Continuous optimization}
Consider a continuous optimization problem with state space
$\scX=\mathbb{R}^n$. The sampling
distribution on $\mathbb{R}^n$ can be quite arbitrary and does not
need to be related to the objective function $S$. Usually, the random vector
$\bX=(X_1,\ldots,X_n)^\top \in \mathbb{R}^n$ is generated from a Gaussian
distribution with independent components, characterized by a vector $\vect{\mu}$ of
means and a vector $\vect{\sigma}$ of standard deviations. At each
iteration of the CE method, these vectors of parameters  are updated as
the means and standard deviation of the elite samples. During the course
of the
algorithm a sequence of $(\vect{\mu_t})$ and $(\vect{\sigma_t})$ are
generated, such that $\vect{\mu_t}$ tends to the optimizer $\bx^*$, while the vector of
standard deviations tends to the zero vector. At the end of the
algorithm one
should obtain a degenerated probability density with mean
$\vect{\mu}_T$ approximately equal to the optimizer $\bx^*$ and all
standard deviations close to 0. A possible stopping criterion is to
stop when all components in $\vect{\sigma}_T$
are smaller than some $\varepsilon$. This scheme is referred to as
{\em normal updating}.

\pkg{CEoptim} implements the normal updating scheme for continuous
optimization.

\subsection{Discrete optimization}
If the state space $\scX$ is finite, the optimization problem is often
referred to as a discrete or combinatorial optimization problem, where
$\scX$ could be the space of combinatorial objects, such as binary
vectors, trees, graphs, etc. To apply the CE method to a discrete
optimization problem, one needs a convenient parameterized
random mechanism to generate samples.

For discrete optimization \pkg{CEoptim} implements sampling from
state spaces $\scX$ of the
form $\{0,1,\ldots,c_1-1\} \times \cdots \times \{0,1,\ldots,c_n-1\}$,
where the $\{c_i\}$ are strictly positive integers. The components of
the random vector $\bX = (X_1,\ldots,X_n)\in \scX$ are taken to be independent, so that its
distribution is determined by a sequence of probability vectors
$\bp_1, \ldots, \bp_n$, with the $j$th component of $\bp_i$
corresponding to $p_{ij} = \Pm(X_i = j)$.
For a given elite sample set
$\scE$ of size $N^{\elite}$, the CE updating formulas for these
probabilities are
\begin{equation}
{p}_{ij}=\frac{\sum_{\bX \in \scE}\I\{X_i = j\} }{N^{\elite}}, \quad i=1,\ldots,n, \quad j=0,\ldots,c_n-1,
\end{equation}
where $\I$ denotes the indicator function. Hence, at each iteration,
probability $p_{ij}$ is updated simply as the average number of times
that the $i$th
component of the elite vectors is equal to $j$.
A possible stopping rule for a discrete optimization problem is to
stop when the overall best objective value does not change over a
number of iterations. Alternatively, one could stop when the sampling
distribution has degenerated sufficiently; for example, when all $\{p_{ij}\}$ are no further than $\epsilon$ away from
either 0 or 1.

\subsection{Constrained optimization}

The general optimization problem (\ref{prob1}) also covers constrained
optimization, where the search space $\scX$ could, for example, be defined by a
system of inequalities:
\begin{equation} \label{eq:constrain}
G_i(\bx)\leq 0, \quad i=1,\ldots,k.
\end{equation}

One way to deal with constraints is
to use \textit{acceptance-rejection}: generate a random vector $\bX$
on a simple search space that contains $\scX$, and accept or reject it based on whether
the sample falls in $\scX$ or not. Alternatively, one could try to
sample
directly from a truncated distribution on $\scX$, e.g., using Gibbs
sampling.

\pkg{CEoptim} implements linear constraints for continuous
optimization of the form $A \bx \leq
\bfb$, where $A$ is a matrix and $\bfb$ a vector. The program will use
either acceptance--rejection or Gibbs sampling to sample from the
multivariate normal distribution truncated to the constraint set.

 A second approach to
handle constraints is to introduce a
\textit{penalty function}. For example, for the constraints
\eqref{eq:constrain}, the objective function could be modified to
\begin{equation}\label{eq:con2}
\tilde{S}(\bx)=S(\bx)+\sum_{i=1}^kH_i\max\{G_i(\bx),0\},
\end{equation}
where $H_i<0$ measures the importance of the $i$th penalty.
 To use the penalty approach with
\pkg{CEoptim} the user simply needs to modify the objective function
according to \eqref{eq:con2}.
The choice of the penalty constants $\{H_i\}$ is problem specific and may need to be
determined by trial and error.

%% \intrusion{Came this far 9/11/2014. The paragraph below needs some
%%   rewriting.}
%%   {\color{red}QB: Our sample method should be described here}\\
%% \pkg{CEoptim} contains an option for truncated sampling for the
%% constrained optimization problems.  therefore, for some constrained
%% problems that cannot be solved by truncated sampling, one should
%% indicate the acceptance-rejection rules in the objective function
%% part, or make the penalty objective function. Then the updating of
%% parameters of sampling distributions can be the same way as for the
%% unconstrained cases.
\section{CEoptim description}

\label{sec:ceusage}
 In this section we describe how to use \pkg{CEoptim}.
%% \intrusion{Please check what packages are required}
%% \begin{itemize}
%% \item The package \pkg{MASS} is required for the function
%%   \code{mvrnorm} to draw from a multivariate normal distribution.
%% \item The package \pkg{msm} is required to *****
%% \item The function \code{rtmvnorm} from \pkg{CEoptim} is required to draw from a
%%   truncated normal distribution.
%% \end{itemize}
% The
%\pkg{CEoptim} is written in \proglang{R} language, based on the
%algorithms in \cite{cebook}.  %
%The use of \pkg{CEoptim} requires a
%functioning installation of the \pkg{CEoptim}. And also, for solving
%the constrained optimization, \pkg{msm} is required, since
%\pkg{CEoptim}  calls \proglang{rtnorm} to ge
%nerate truncated samples.

The \code{CEoptim} function is the main function of the package
\pkg{CEoptim}. It can be used to  solve continuous and discrete
optimization
problems as well as mixtures thereof.
\newpage
\subsection{Usage}

\begin{verbatim}
CEoptim(f, f.arg=NULL, maximize=FALSE, continuous=NULL, discrete=NULL,
        N=100L, rho=0.1, iterThr=1e4L, noImproveThr= 5, verbose=FALSE)
\end{verbatim}
%\intrusion{Make it as close as possible to the help file}

\subsection{Arguments}
%\intrusion{Make sure all arguments are discussed below}
%\intrusion{Needs to be able to span over different pages} Done

%\begin{longtable}{ p{.15\textwidth} p{.1\textwidth} p{.75\textwidth} }
\begin{longtable}{ p{.20\textwidth}   p{.80\textwidth} }
\addtocounter{table}{-1} 
\bf Argument  & \bf Description \\[2pt]
\hline \\
\code{f} &   Function to be optimized. Can have continuous
  and discrete arguments.\\[10pt]

\code{f.arg} &   List of additional fixed arguments passed to function
  \code{f}. \\[10pt]

  \code{maximize} & Logical value determining whether to maximize or minimize the objective function. \\[10pt]

\code{continuous} &  List of arguments for the continuous
optimization part, consisting of:\\[4pt]
 --- \quad \code{mean} & Vector of initial means.\\[4pt]
--- \quad \code{sd} & Vector of initial standard deviations.\\[4pt]
%\code{bounds} & A $2\times p$ numeric \textbf{matrix} which defines the ranges of the continuous arguments. The $i$-th continuous variable is constrained to lie in the interval (\code{bounds}$[1,i]$,\code{bounds}$[2,i]$).\\[4pt]
--- \quad \code{smoothMean} & Smoothing parameter for the
vector of means.  Default
 value 1 (no smoothing).\\[4pt]
--- \quad \code{smoothSd} &  Smoothing parameter for the standard
deviations. Default value 1 (no smoothing).\\[4pt]
--- \quad \code{sdThr} & Positive numeric convergence threshold.  Check whether
 the maximum standard deviation
is smaller than \code{sdThr}. Default value 0.001.\\[4pt]
--- \quad \code{conMat} & Coefficient matrix of linear constraint
\code{conMat} $\bx \leq $ \code{conVec}. \\[4pt]
--- \quad \code{conVec} & Value vector of linear constraint  linear constraint
\code{conMat} $\bx \leq $  \code{conVec}.   \\[10pt]

\code{discrete} &  List of arguments for the discrete
optimization part, consisting of:\\[4pt]
--- \quad \code{categories} & Integer vector which defines the allowed
values of the categorical variables. The \code{i}th categorical
variable takes values in the set $\{0,1,\ldots,\text{\code{categories(i)}}-1\}$.
\\[4pt]
--- \quad \code{probs} & List of  initial probabilities for the
categorical variables. Defaults to equal (uniform) probabilities.\\[4pt]

--- \quad \code{smoothProb} & Smoothing parameter for the probabilities of
the categorical sampling distribution. Default
 value 1 (no smoothing).\\[4pt]
--- \quad \code{probThr} & Positive numeric convergence
threshold. Check whether all probabilities
in the categorical sampling distributions
deviate less than \code{probThr} from either 0 or 1. Default value 0.001.\\[10pt]

\code{N} & Integer representing the CE sample size. \\[4pt]
\code{rho} & Value between 0 and 1 representing the elite
proportion. \\[4pt]

\code{iterThr} & Termination threshold on the
largest number of iterations. \\[4pt]
\code{noImproveThr} & Termination threshold on the largest
number of iterations during which no improvement of the best function
value is found. \\[4pt]

\code{verbose} & Logical value set for CE progress
output. \\[4pt]

\end{longtable}

%\intrusion{We should be able to plot the
%  progress after a verbose session where the relevant vectors are stored.}
\subsection{Value}
\code{CEoptim} returns a list with the following components.
\begin{longtable}{ p{.20\textwidth}  p{.80\textwidth} }
\addtocounter{table}{-1} 
\code{optimum} & Optimal value of \code{f}. \\[10pt]
\code{optimizer} &  List of the location of optimal value, consisting
of:\\[4pt]
--- \quad \code{continuous} & Continuous part of the optimizer. \\[4pt]
 --- \quad \code{discrete} &  Discrete part of the optimizer.\\[10pt]
\code{termination} & List of termination information consisting of:\\[4pt]
--- \quad \code{niter} & Total number of iterations upon
termination. \\[4pt]
--- \quad \code{convergence} & One of the following  termination statements:\vspace{-5pt}
\begin{itemize}
\setlength{\itemsep}{-5pt}
\item \code{Not converged}, if the number of iterations reaches \code{iterThr};
\item  \code{The optimum did not change for noImproveThr iterations}, if the
  best value has not improved for \code{noImproveThr} iterations;
\item \code{Variances converged}, otherwise.
\end{itemize} \\
\code{states} & List of intermediate results computed at each
iteration. % Will only be returned when \code{verbose=TRUE}.
 It consists of the iteration number (\code{iter}), the best overall
 value
 (\code{optimum}) and the worst value of the elite
 samples, (\code{gammat}). The means (\code{mean}) and maximum standard deviation
 (\code{maxSd}) of the elite set are also included for continuous
 cases, and the maximum deviations (\code{maxProbs}) of the sampling probabilities to either $0$ or $1$ are included for discrete cases. \\
 \code{states.probs} & List of categorical sampling probabilities
 computed at each iteration. Will only be returned for discrete and mixed cases.
\end{longtable}

%\intrusion{The states variable needs further specification. What does
%  it consist of in the continuous, discrete and mixed case.}
%\intrusion{We need to specify what termination information}
\subsection{Note}
\begin{itemize}
\item Although partial parameter passing is allowed outside lists, it is
  recommended that parameters names are specified in full. Parameters
  inside lists have to specified completely.
\item Because \code{CEoptim} is a random function it is useful to (1)
  set the seed for the random number generator (for testing purposes),
  and (2)
  investigate the quality of the results by repeating
  the optimization a number of times.
\end{itemize}
\newpage
\section{Numerical examples}\label{sec:numerical}
The following examples illustrate the use, flexibility, and efficacy of the
\code{CEoptim} function from the package \pkg{CEoptim}.

\subsection{Maximizing the peaks function}\label{eg:peak}
\rm
Suppose we wish to maximize $\matlab$'s well-known {\em peaks} function, given by
\begin{equation}\label{peak}
S(\bx)=3(1-x_1)^2\, \e^{-x_1^2-(x_2+1)^2}-10\left(\frac{x_1}{5}-x_1^3-x_2^5\right)\e^{-x_1^2-x_2^2}-\frac{1}{3}\e^{-(x_1+1)^2-x_2^2}\;.
\end{equation}
\begin{figure}[H]
%source: ./Programs/peaksplot.R
\centerline{\includegraphics[width=0.5\linewidth]{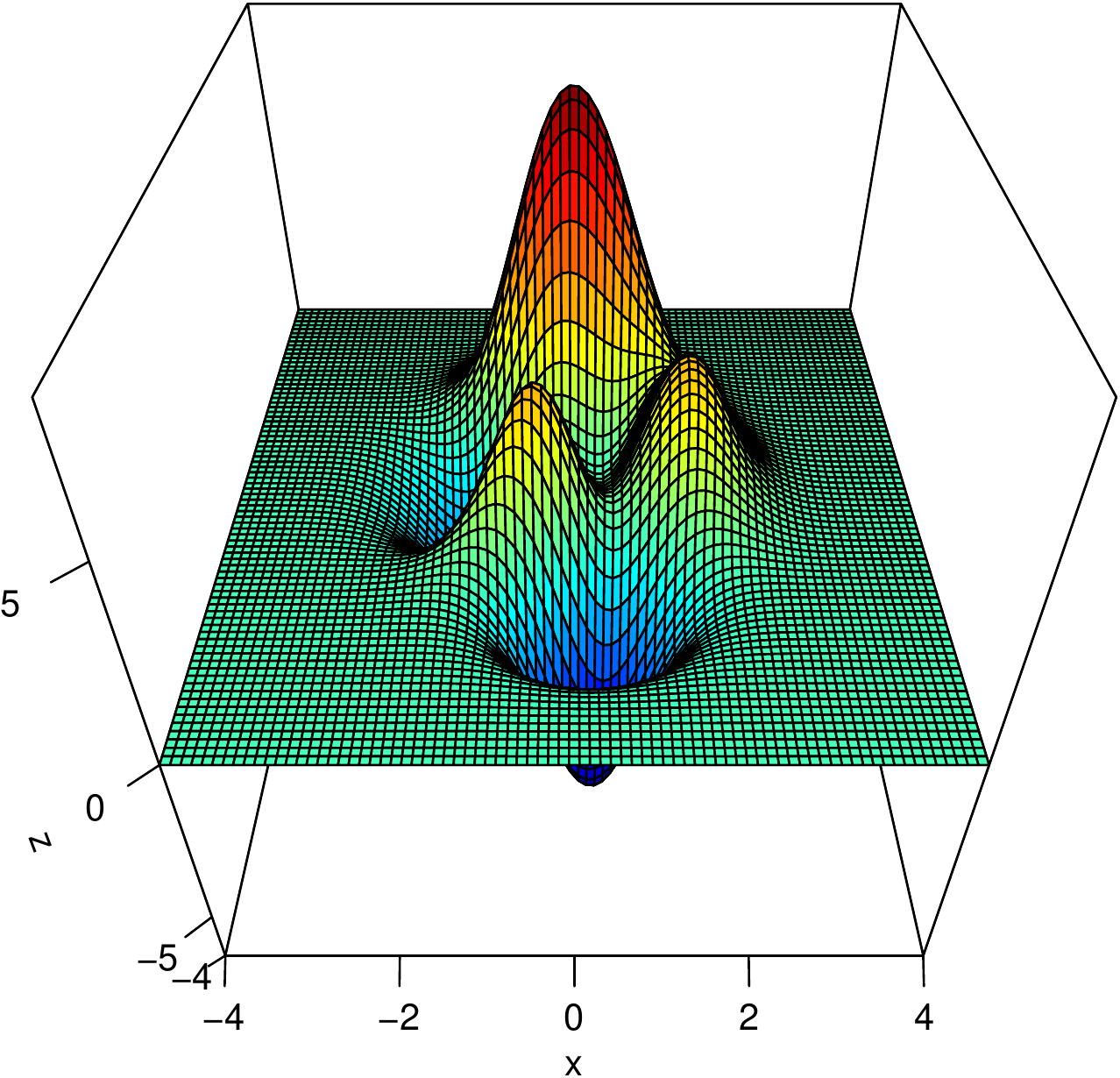}}
\caption{Peaks function}
\end{figure}
%\intrusion{Remove whitespace around figures. In the latex file show
%  where the program can be found that made the figure. Do this for all figures}
The peaks function has three local maxima and three local
minima, with a global maximum at $\bx^*\approx(-0.0093,1.58)$ of
 $S(\bx^*)\approx 8.1$, and the other two local maximum are $S(\bx_1)=3.78$ at $(-0.46,-0.63)$ and $S(\bx_2)=3.59$  at $(1.29,-0.0049)$.

%\intrusion{optim doesn't work for starting value (0,0)}
% 3.776581 at (-0.4600405,-0.6292215)
% 3.59249  at (1.285678949,-0.004894594)

To solve the problem with \code{CEoptim}, using normal updating, we
must specify the vector of initial means $\bmu_0$ and standard
deviations $\vect{\sigma}_0$ of the
2-dimensional Gaussian sampling distribution. The initial sampling
distribution should cover, roughly, the region where the maximizer is thought to
lie. As an example we take $\bmu_0=(-3,-3)$ and
$\vect{\sigma}_0=(10,10)$. The important point is that the standard
deviations are chosen large enough.
Since this is a
maximization problem, we have to set \code{maximize=T}.
For the other parameters
we take their default values. Note that there are only four parameters
to be updated in each iteration, so a sample size of $N=100$ is suitable.
%\intrusion{Putting the + at the beginning of the line is not OK.}
%source: ./Programs/example-peaks.R
\begin{CodeInput}
R> require(CEoptim)
R> fun <- function(x){3*(1-x[1])^2*exp(-x[1]^2 - (x[2]+1)^2)-10*(x[1]/5 +
                      -x[1]^3 - x[2]^5)*exp(-x[1]^2 - x[2]^2) +
                      -1/3*exp(-(x[1]+1)^2 - x[2]^2)}

R> set.seed(1234)  # for verification purpose only
R> mu0 <- c(-3,-3); sigma0 <- c(10,10)
R> res <- CEoptim(fun, maximize=T, continuous=list(mean=mu0,sd=sigma0))
R> res
\end{CodeInput}

The output of this implementation is as below:
\begin{CodeOutput}
Optimizer for continuous part: 
  -0.009390034 1.581405 
Optimum: 
  8.106214 
Number of iterations: 
  7 
Convergence: 
  Variance converged 
\end{CodeOutput}

The reader may check that \code{optim} applied to the minimization of
$-f$ can easily find the wrong optimizer, e.g., when the starting
value is $(0,0)$.

\subsection{Non-linear regression}\label{eg:mf}
\rm

\begin{figure}[htb]
\centerline{\includegraphics[width=0.8\linewidth]{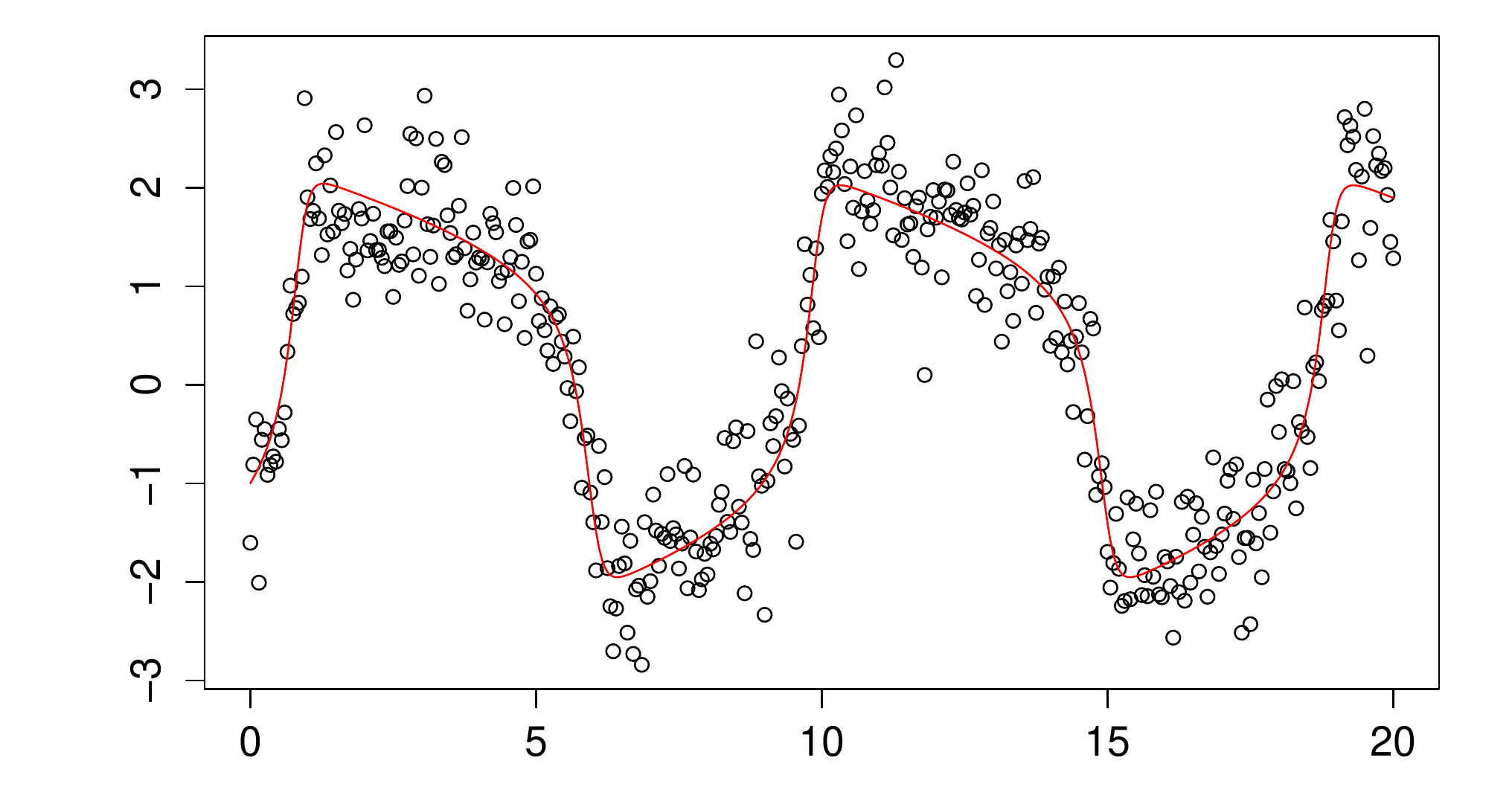}}
\caption{Simulated data (points) and ``unknown'' true
curve (red).}
\label{fig:contopt}
\end{figure}

We next consider a more complicated optimization task, involving data
generated from the well-known {\em FitzHugh--Nagumo} differential equations:
\begin{equation}\label{eq:fitznag}
\begin{split}
\frac{\di V_t}{\di t} & = c \left( V_t - \frac{V_t^3}{3} + R_t\right)\;,\\
\frac{\di R_t}{\di t} & = - \frac{1}{c} (V_t - a + b R_t)\;,
\end{split}
\end{equation}
which model the behavior of certain types of neurons \citep{nagumo62}.
\cite{Ramsay07} consider estimating the parameters $a$, $b$, and $c$
from noisy observations of $(V_t)$
by using a generalized smoothing approach. The simulated data
 in
Figure~\ref{fig:contopt} (saved as \code{data(FitzHugh)})
correspond to the values of $V_t$ obtained from
\eqref{eq:fitznag} at times
$0,0.05,\ldots,20.0$,  adding Gaussian noise with standard
deviation 0.5. That is, we use the non-linear regression model
\begin{equation}\label{mod:Fitz}
Y_i = V_{0.05 i}(\bx) + \epsilon_i, \quad i = 1,\ldots,400\;,
\end{equation}
where the
$\{\epsilon_i\}$
are iid with a $\Nor(0,\sigma^2)$ distribution,
$V_{0.05 i}(\bx)$ is the solution to
\eqref{eq:fitznag} for time $t = 0.05 i$, and  $\bx = (a,b,c,V_0,R_0)$
is the vector of parameters.
 The true parameter values are here $a = 0.2$, $b=0.2$, and
$c =3$. The initial conditions are $V_0 = -1$ and $R_0 = 1$.

% figure and data made with ./Programs/Fitzhughplot.R
Estimation of the parameters via the CE method can be established by
minimizing the least-squares performance
\begin{equation}\label{eq:mfobj}
S(\bx) = \sum_{i=0}^{400} \left(y_{i} - V_{0.05 i}(\bx) \right)^2\;,
\end{equation}
where the $\{y_i\}$ are the simulated data from the model
\eqref{mod:Fitz}.
%,  for parameter vector $\bx$.
Note that we assume
that also the initial conditions are unknown.
%\begin{figure}[H]
%\centerline{\includegraphics[width=0.95\linewidth]{contopt}}
%\caption{Simulated data for the FitzHugh--Nagumo model and a fitted
%curve obtained via the CE method.}
%\label{fig:contopt}
%\end{figure}

%% \intrusion{
%% I think this is old code.
%% Please modify the code to exactly reflect the correct usage of
%% CEoptim. For one thing, the function FN needs to be taken out of the
%% loop. And we need to parse the parameters via the ... construction.}

%% , which is to solve the differential equations
%% numerically for different parameters and initial values input. The
%% output of \code{ssres} is $S(\bx)$ shown in Equation
%% \ref{eq:mfobj}.

We use the \pkg{deSolve} package to numerically solve the
FitzHugh--Nagumo differential equations \eqref{eq:fitznag}.
Hereto, we first define the function \code{FN}.
\newpage

\begin{CodeInput}
R> FN <- function(t,state,parameters){
       with(as.list(c(state,parameters)),{
       dV <- c*(V-V^3/3+R)
       dR <- -1/c*(V-a+b*R)
       list(c(dV,dR))
      })}
\end{CodeInput}

The following function \code{ssres} now implements
the objective function in \eqref{eq:mfobj}.

\begin{CodeInput}
R> ssres <- function(x,fundf,times,y) {
  		parameters <- c(a=x[1],b=x[2],c=x[3])
  		state <- c(V=x[4],R=x[5])
  		out <- ode(y=state,times=times,func=fundf,parms=parameters)
  		return(sum((out[,2]-y)^2))}
\end{CodeInput}
\code{CEoptim} could be used with ${\bmu}_0=(0,0,5,0,0)$ and ${\vect{\sigma}}_0=(1,1,1,1,1)$. Constant smoothing
parameters $\alpha = 0.9$ and $\beta= 0.5$ were used for the
$\{{\bmu}_t\}$ and the $\{{\vect{\sigma}}_t\}$, respectively. To see
the progress of the algorithm we set \code{verbose} to
\code{TRUE}. The other
arguments  remain default.

%code: ./Programs/example-fitzhugh.R
\begin{CodeInput}
R> require(deSolve)
R> require(CEoptim)
R> set.seed(123405)
R> times <- seq(0,20,by=0.05)
R> data(FitzHugh)

R> res<- CEoptim(ssres, f.par = list(fundf=FN, times=times, y=ySim),
                 continuous= list(mean=c(0,0,5,0,0), sd=c(1,1,1,1,1),
                 smoothMean=0.9,smoothSd=0.5), verbose=TRUE)
\end{CodeInput}
%\intrusion{This is not correct any more regarding the output information}
The final output is as follows:
\begin{CodeOutput}
R> res
Optimizer for continuous part: 
 0.1959748 0.2395983 3.001453 -0.9938222 0.9791585 
Optimum: 
 102.8005 
Number of iterations: 
 41 
Convergence: 
 Variance converged 
\end{CodeOutput}
%\intrusion{What do we mean with convergence=TRUE? we need to unify the
%convergence information}
%\intrusion{This already is stated above. so needs to be linked
%  better. The output show the estimates ... The MLE of
%  $\sigma^2$ is given by optimum/400}
The output shows the estimates (notice that the initial condition
was assumed to be unknown):
${\hat{a}} = 0.1959748, {\hat{b}} = 0.2395983, {\hat{c}} =3.0014526, {\hat{V}}_0 = -0.9938222$,
and ${\hat{R}}_0 =  0.9791585$, with the maximum likelihood estimate
$\hat{\sigma} = \sqrt{102.8005/400} = 0.507$  for the residual
standard deviation $\sigma$. The reader may check that fitted curve
is practically indistinguishable from the true one in Figure~\ref{fig:contopt}.

\begin{figure}[htb]
\centerline{\includegraphics[width=0.8\linewidth]{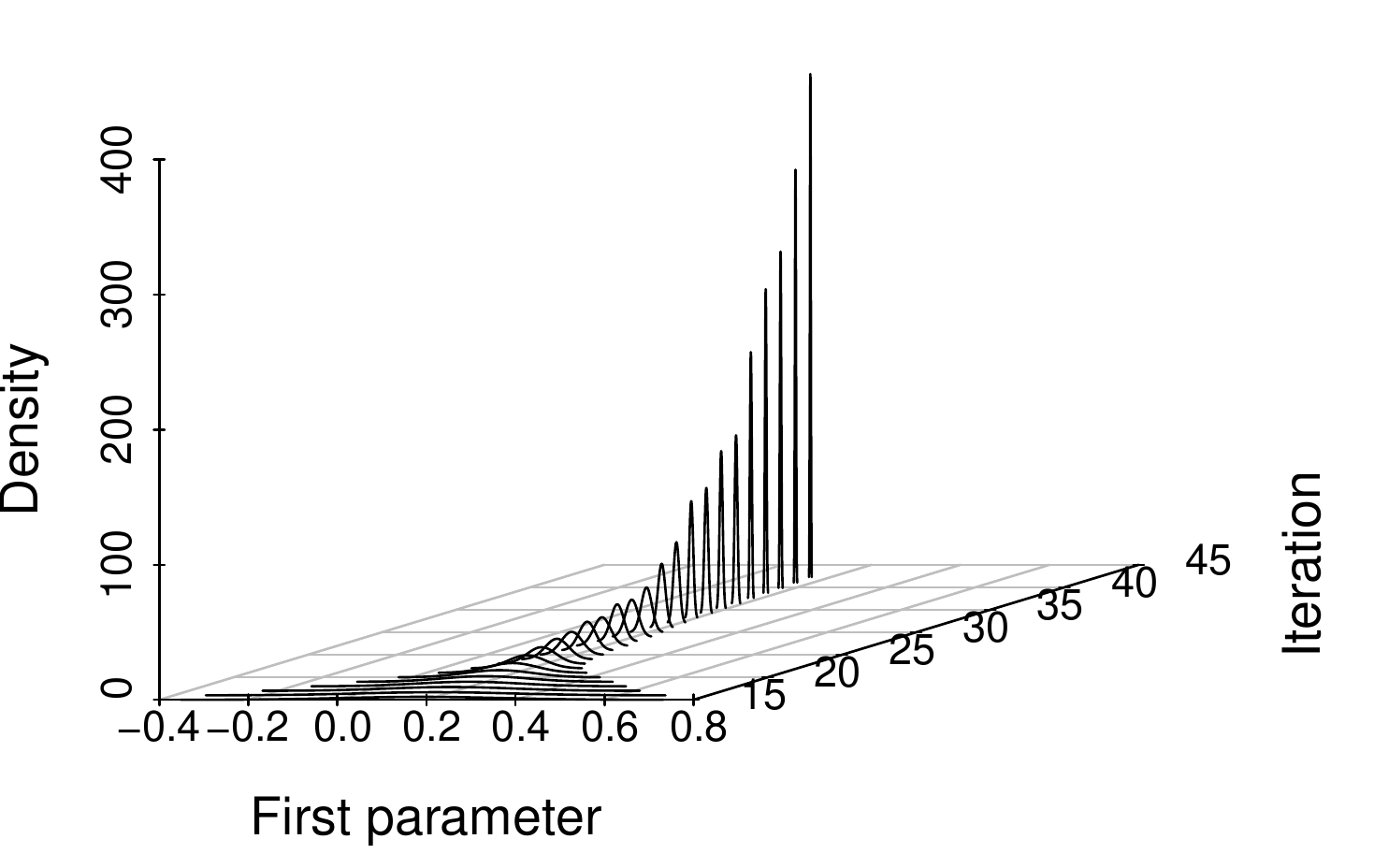}}
\caption{The evolution of the sampling pdf for the first parameter $a$}
\label{fig:pdfevolution}
\end{figure}

To illustrate how the sampling distributions change during the CE process,
we have plotted in Figure~\ref{fig:pdfevolution} the evolution of the
sampling pdf for the first parameter $a$,
from the 15th to the final iteration. As can be seen from the figure,
the sampling distribution converges to a point distribution around the
optimal value for $a$.

\newpage

\subsection{Max-cut problem}\label{eg:maxcut}
\rm
The max-cut  problem in graph theory can be formulated as follows.
Given a weighted graph $(V,E)$ with node set $V = \{1,\ldots,n\}$ and edge set $E$,
partition the nodes of the graph into two  subsets
$V_1$ and $V_2$
 such that
the sum of the (nonnegative) weights of the edges going from one subset to the
other is maximized. Let $C = (C_{ij})$ be the matrix of weights. The objective is to maximize
\begin{equation}\label{cutval}
\sum_{(i,j) \in V_1 \times V_2} (C_{ij} + C_{ji})\;
\end{equation}
over all {\em cuts}  $\{V_1,V_2\}$.
Such a cut can be conveniently represented by a binary {\em cut
vector} $\bx = (x_1,x_2,\ldots,x_n)$, where $x_i = 1$ indicates that
 $i\in V_1$. Let $\scX$ be the set of cut vectors and let $S(\bx)$ be the
value of the cut represented by $\bx$, as given
in \eqref{cutval}.

To maximize $S$ via the CE method one can generate the random cut
vectors by drawing each component (except the first one, which is set
to 1) independently
from a Bernoulli distribution, that is, $\bX =
(X_1,X_2,\ldots,X_n) \sim \ber(\bp)$, where
 $\bp =(1,p_2,\ldots,p_n)$.
%Given an elite sample set $\scE$, with size $N^\e$,  the updating
In this case the updated success probability for the $i$th component is
the mean of the $i$-th components of the vectors in the elite set.

% Figure~\ref{fig:maxcut} illustrates the evolution of the
%Bernoulli parameters for a max-cut problem from \cite{de2005tutorial} of
%dimension $n=400$,
% for
%which the optimal solution is given by $\bx^* =(1,\ldots,1,0,\ldots,0)$.

%\begin{figure}
%\begin{center}
%\includegraphics[width=8cm]{FIG/ESS_data.pdf}
%\caption{\label{fig-data} General representation of data for the
%  investigation of OMICs biomarkers.}
%\end{center}
%\end{figure}

%\intrusion{Let's make this a more practical problem. I'd be nice if we
%could relate the matrix to some real data, e.g., human relationships
%(who knows who).}

\begin{figure}[htb]
\centerline{\includegraphics[width=0.6\linewidth]{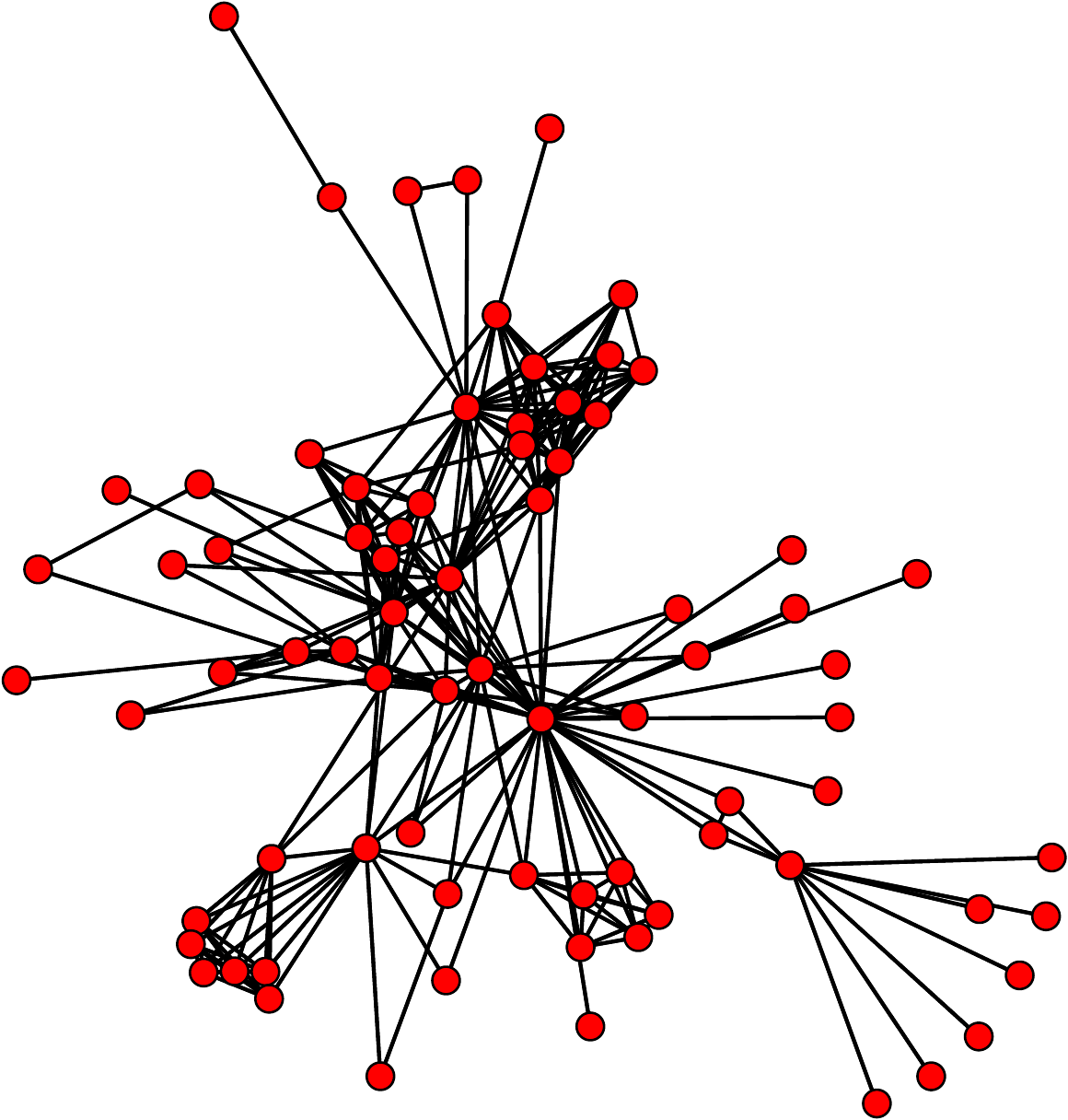}}
\caption{Network of coappearances}
\label{fig:network}
\end{figure}

%gephi has lesmis as a standard data set.
%% See also https://wiki.gephi.org/index.php?title=Datasets
As an example, consider the network from \cite{lesmis_knuth} describing
the coappearances of 77 characters from Victor Hugo's novel {\em Les
  Miserables}. Each node of the network represents a selected
character and edges connect any pair of characters that coappear.
The weights of the edges are the number of such coappearances. Using
\pkg{CEoptim}, the
data can be loaded via the command \code{data(lesmis)}. The network is
displayed in Figure~\ref{fig:network}, using the graph
analysis
package \pkg{sna}.
\begin{CodeInput}
R> library(sna)
R> library(CEoptim)
R> data(lesmis)
R> gplot(lesmis,gmode="graph")
\end{CodeInput}

For any fixed cost matrix \code{costs} and cut vector \code{x}, the objective function of the max-cut problem can be written as:
\begin{CodeInput}
R> fmaxcut <- function(x,costs){
      v1 <- which(x==1)
      v2 <- which(x==0)
      return( sum(costs[v1,v2])) }
\end{CodeInput}

\begin{figure}[htb]
\centerline{\includegraphics[width=\linewidth]{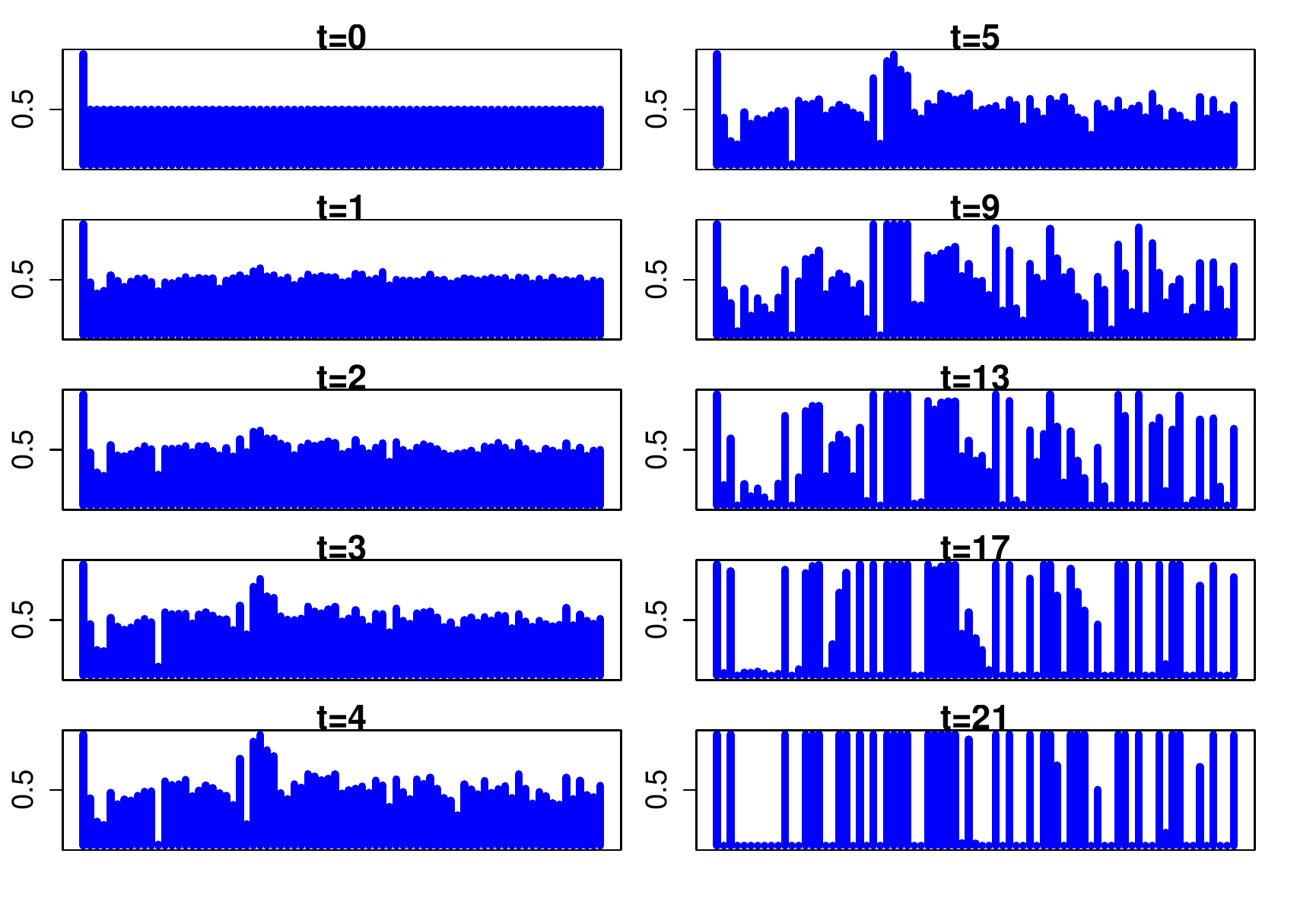}}
\caption{Evolution of categorical sampling probabilities that characters in group 1.}
\label{fig:probsevolution}
\end{figure}

To optimize this function  with the \pkg{CEoptim} package, we specify the
following arguments:  \code{discrete\$probs=\{(0,1);} \code{(0.5.0.5);\ldots ;(0.5,0.5)\}}, sample size \code{N=3000} and
optimization type: \code{maximize=T}. To see the output we set
\code{verbose=TRUE}.
The other arguments are taken as default.
Note that users only need to specify either \code{categories} or \code{probs}, if  both of them are specified, then \code{categories} will be overridden.
\begin{CodeInput}
R> set.seed(5)
R> p0<-list()
R> for(i in 1:77){p0<-c(p0,list(rep(0.5,2)))}
R> p0[[1]] = c(0,1)
R> res <- CEoptim(fmaxcut,f.arg=list(costs=lesmis),maximize=T,
          verbose=TRUE,discrete=list(probs=p0),N=3000L)

R> ind <- res$optimizer$discrete
R> group1 <- colnames(lesmis)[which(ind==TRUE)]
R> group2 <- colnames(lesmis)[which(ind==FALSE)]
\end{CodeInput}
The output of \code{CEoptim} is as follows:
%\intrusion{Again, this is no longer accurate due to the changed termination info}
\begin{CodeOutput}
R> res
Optimizer for discrete part: 
 1 0 1 0 0 0 0 0 0 0 1 0 0 1 1 1 0 0 1 1 0 1 0 
 1 0 1 1 1 1 0 0 1 1 1 1 1 1 0 0 0 0 1 0 1 0 0 
 1 0 1 1 1 0 1 1 1 0 0 0 0 1 1 0 1 0 0 1 1 1 1 
 0 0 1 0 1 0 0 0 
Optimum: 
 535 
Number of iterations: 
 20 
Convergence: 
 Optimum did not change for 5 iterations 
\end{CodeOutput}
Note that character 1 (Myriel) is always in \code{group1}. The initial
probabilities for the other characters are $0.5$. With \code{states.probs}, we can plot the
evolution of the probabilities that each character belongs to \code{group1}; see Figure \ref{fig:probsevolution}.
\begin{CodeInput}
R> probs <- res$states.probs
R> X <- matrix(NA,nrow=length(probs),ncol=77)
R> prob0 <- cbind(1,t(rep(0.5,76)))
R> for(i in 1:length(probs)){
       for(j in 1:77){
           X[i,j] <- res$states.probs[[i]][[j]][2]
        }
      }
R> X <- rbind(prob0,X)
R> par(mfcol=c(5,2),mar=c(1,1.5,1,1.5),oma=c(1,1,1,1))
R> for(i in 1:5){
      plot(X[i,],type="h",lwd=4,col="blue",ylim=c(0,1),xaxt="n",yaxt="n",ylab="",
      main=paste("t=",i-1,sep=""))
      axis(2,at=0.5,labels=0.5)}
R> for(i in 1:5){
      plot(X[1+4*i,],type="h",lwd=4,col="blue",ylim=c(0,1),xaxt="n",yaxt="n",ylab="",
      main=paste("t=",1+4*i,sep=""))
      axis(2,at=0.5,labels=0.5)}
\end{CodeInput}

\begin{figure}[H]
\centerline{\includegraphics[width=0.8\linewidth]{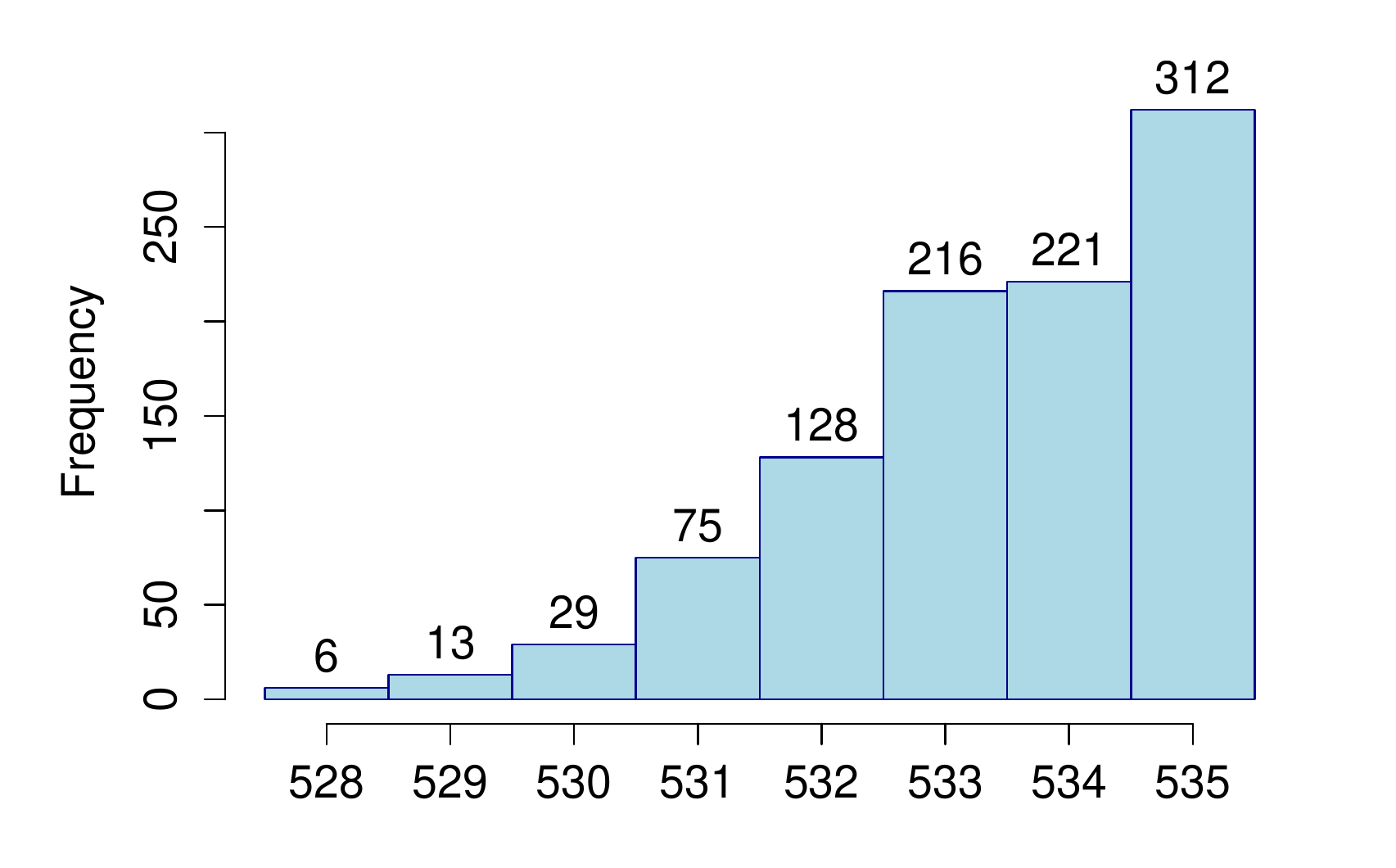}}
\caption{Frequency of best max-cut values found by \code{CEoptim}.}
\label{fig:hist}
\end{figure}

Based on the output above, the two groups of characters are indicated in Table \ref{tab:groups}:
%\intrusion{We need to show the new termination condition}
\begin{table}[H]
\centering
\begin{longtable}{ p{.5\textwidth} | p{.5\textwidth} }
 \code{group1}  &   \code{group2}  \\[3pt]
\hline
\\[-7pt]
\code{Myriel,             MlleBaptistine,     Labarre,            MmeDeR,
  Isabeau,            Gervais,            Fameuil,            Blacheville ,
  Dahlia,             Fantine,            Thenardier,         Cosette,
  Javert,             Fauchelevent,       Simplice,           Scaufflaire,
  Oldwoman1,          Judge,              Champmathieu,       Brevet,
  Eponine,            Oldwoman2,          Jondrette,          Gavroche,
  Gillenormand,       Magnon,             MmePontmercy,       MlleVaubois,
  LtGillenormand,     Combeferre,         Prouvaire,          Courfeyrac,
  Joly,               Grantaire,          MotherPlutarch,     Gueulemer,
  Montparnasse,       Child1        }

 &\code{Napoleon, MmeMagloire, CountessDeLo, Geborand,
 Champtercier, Cravatte, Count, OldMan,
  Valjean, Marguerite, Tholomyes,  Listolier,
  Favourite,          Zephine,            MmeThenardier,      Bamatabois,
  Perpetue,           Chenildieu,         Cochepaille,        Pontmercy,
  Boulatruelle,       Anzelma,            MotherInnocent,     Gribier,
  MmeBurgon,          MlleGillenormand,   Marius,             BaronessT,
  Mabeuf,             Enjolras,           Feuilly,           Bahorel,
  Bossuet,            Babet,              Claquesous,         Toussaint,
  Child2,             Brujon,            MmeHucheloup
  }     \\
\caption{Two groups of characters providing a maximal cut.}
\label{tab:groups}
\end{longtable}

\end{table}
% done \intrusion{Check if the names correspond to Knuth's original in
%  jean.dat. The two Woman should probably be Woman1 and Woman2.}
We have run the program for 1000 times randomly. In  312 cases the
optimal solution (535) was found. The frequency of the
results of \code{CEoptim} is given in Figure \ref{fig:hist}.
%\intrusion{Remove the heading from the figure}

\newpage
\subsection{Constrained minimization of the griewank function}\label{eg:Griewank}

To illustrate constrained optimization with \code{CEoptim}, we
consider the minimization of the {\em griewank} function, which is
widely used to test the convergence of
optimization algorithms. The griewank function of order $n$ is defined as
\begin{equation}\label{griewank}
S(\bx)=1+\frac{1}{4000}\sum_{i=1}^nx_i^2-\prod_{i=1}^n\cos\left(\frac{x_i}{\sqrt{i}}\right),
\end{equation}
where $\bx =(x_1,\ldots,x_n)^\top$ takes values in some subset of
$\R^n$.
The function has many local minima with (in the unconstrained case) a
global minimum at $\bx^*=(0,\ldots,0)$ of $S(\bx^*)=0$.

\begin{figure}[H]
\centerline{\includegraphics[width=0.9\linewidth]{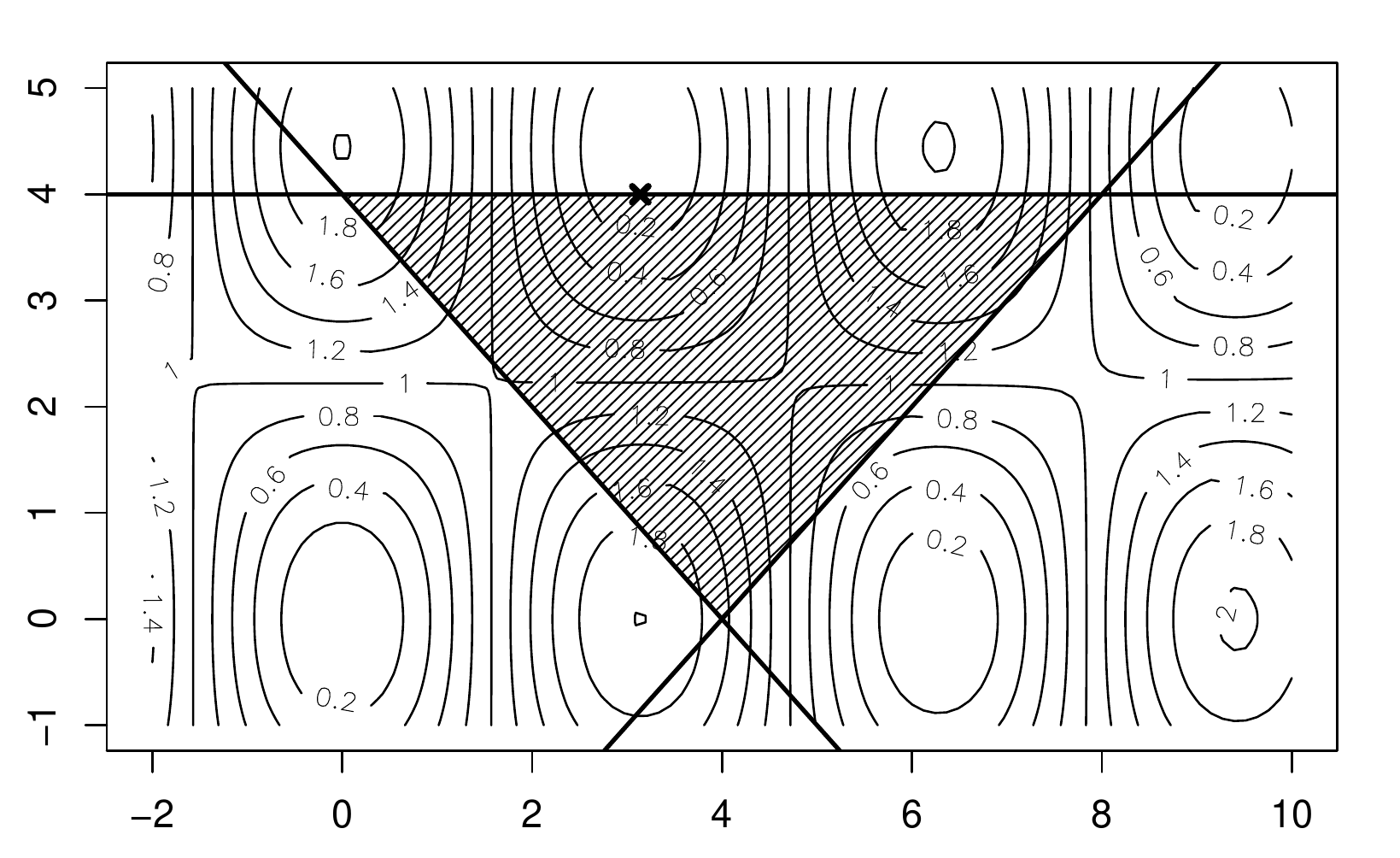}}
\caption{Contour plot of the griewank function and the triangular
  constraint region.
 The optimal
  solution (indicated by a cross) lies on the boundary of the
  constraint region.}
\label{fig:griewankcontour}
\end{figure}

We wish to minimize the griewank function of order 2 over the triangle
with vertex points $(1,4)$, $(4,0)$, and $(8,4)$; see
Figure~\ref{fig:griewankcontour}.

%\intrusion{Make the axes font a bit smaller}

The constraint set can be written
as the linearly constrained region $\{\bx \in \R^2: A \bx \leq \bfb\}$ with
\[
A  = \begin{pmatrix}
0 & 1 \\
-1 & - 1 \\
1 & 1
\end{pmatrix}
\text{ and }
\bfb = \begin{pmatrix}
4 \\
-4 \\
4
\end{pmatrix}.
\]

To solve the problem with \code{CEoptim} we proceed as follows:

\begin{CodeInput}
R> require(CEoptim)
R> set.seed(123)
R> griewank <- function(X) {
     p <- length(X)
     r <- c()
     for (i in 1:p) {
       r[i] <- cos(X[i]/sqrt(i))
     }
     return(1+sum(X^2)/4000-prod(r))
   }
R> A <- rbind(c(0,1),c(-1,-1),c(1,-1))
R> b <- c(4,-4,4)
R> res <- CEoptim(griewank,continuous=list(mean=c(0,0), sd=c(10,10), conMat=A,
                  conVec=b), rho=0.1, N=200L, verbose=TRUE, noImproveThr=Inf)

R> cat("direct optimizer =", res$optimizer$continuous,"\n")
R> cat("direct minimum =",res$optimum,"\n")
\end{CodeInput}

The corresponding output shows that the minimum is obtained at the
boundary of the triangle.
\begin{CodeOutput}
R> direct minimizer = 3.139669 3.991955
R> direct minimum = 0.05685487
\end{CodeOutput}

It is also possible to use a penalty approach for this problem.
Here we take the penalty function
\[
\widetilde{S}(\bx) = S(\bx) + 100 \, \| A \bx - \bfb\|,
\]
which can be implemented in the following way.
%\intrusion{Should we call griewank instead inside the function?}
\begin{CodeInput}
R> griewank.penalty <- function(X,A,b) {
     fn <- griewank(X)
     if (any(A%*% as.vector(X) > b)){
        penalty <-  norm(A%*% as.vector(X)- b)
        fn <- fn + 100*penalty}
     return(fn)
   }
\end{CodeInput}

The optimization now proceeds as follows (note that we have also changed
\code{rho} and \code{N}):
\begin{CodeInput}
R> set.seed(123)
R> res.pen<- CEoptim(griewank.penalty,f.arg=list(A,b),continuous=list(mean=c(0,0),
              sd=c(10,10)),rho=0.01,N=2000L,verbose=TRUE,noImproveThr=Inf)

R> cat("penalty minimizer =",res.pen$optimizer$continuous,"\n")
R> cat("penalty minimum =",griewank(res.pen$optimizer$continuous),"\n")
\end{CodeInput}
This leads to practically the same result:
\begin{CodeOutput}
R> penalty minimizer = 3.139757 4
R> penalty minimum = 0.055103
\end{CodeOutput}

\subsection{Dirichlet data} \label{eg:mle}
\rm
Suppose that we are given a random sample of data from a
$\Dirichlet(\balpha)$ distribution,
where $\balpha =(\alpha_1,\dots,\alpha_{K+1})^\top$ is an unknown parameter
vector satisfying $\alpha_i>0$, $i=1,\dots,K+1$. Recall that the pdf
of a random vector
$\bY = (Y_1,\ldots,Y_K) \sim \Dirichlet(\balpha)$ is given by
\[
f(\by; \vect{\alpha}) = \frac{\Gamma(\sum_{i=1}^{K+1}
  \alpha_i)}{\prod_{i=1}^{K+1} \Gamma(\alpha_i)} \prod_{i=1}^K
y_i^{\alpha_i - 1} \left( 1 - \sum_{i=1}^K y_i\right)^{\alpha_{K+1} - 1},
\]
for $x_i \geq 0, i=1,\ldots,K$ and $\sum_{i=1}^K y_i \leq 1$, where
$\Gamma$ is the gamma function.
The conditions on $\balpha$ provide natural inequality constraints:
$G_i(\balpha) \equiv -\alpha_i  \leq 0$, $i=1,\dots,K+1$.

We will use CE  method to
obtain the maximum likelihood estimate by direct maximization of the log-likelihood for the Dirichlet
distribution given the data.

For a particular example, a data size of $n=100$ points are sampled
from the $\Dirichlet(1,2,3,4,5)$ distribution with the assistance of
the function \code{rdirichlet} in the \pkg{CEoptim} package. 

\begin{CodeInput}
R> require(CEoptim)
R> set.seed(12345)
R> a <- 1:5
R> K <- length(a)-1
R> n <- 100
R> y <- dirichletrnd(a,n)
\end{CodeInput}
To use \pkg{CEoptim} to solve the MLE problem.
The objective function is written as follows:
\begin{CodeInput}
R> dirichletLoglike <- function(alpha,Y,n,K){
        t <- apply(Y,MARGIN=1,function(y){sum((alpha[1:K]-1)*log(y[1:K]))+
            (alpha[K+1]-1)*log(1-sum(y[1:K]))})
        out <- n*(log(gamma(sum(alpha)))-sum(log(gamma(alpha))))+sum(t)
        return(out)}
\end{CodeInput}
 The CE parameters
are initial mean vector $\bmu=(0,0,0,0,0)$ and standard deviation vector $\vect{\sigma}=(10,10,10,10,10)$. The sample size of $N=10^4$ and the elite ratio is default. To pass the linear constraints that $\alpha_i>0,i=1,\ldots,K+1$, the  coefficient matrix is
\[A=\begin{pmatrix}
-1 &0& 0& 0& 0 \\
0 &-1& 0& 0& 0 \\
0 &0& -1& 0& 0 \\
0 &0& 0& -1& 0 \\
0 &0& 0& 0& -1 \\
\end{pmatrix},
\]
and the constraint  vector is $\bfb=(0,0,0,0,0)$.
No smoothing parameter is applied to the mean vector, but a constant smoothing
parameter of \code{smoothSd=0.5} is applied to each of the standard deviations. This is a maximization problem, so set \code{maximize=T}.
\begin{CodeInput}
R> mu0 <- rep(0,times=K+1)
R> sigma0 <- rep(10,times=K+1)
R> A <- matrix(rep(0,times=25),nrow=5)
R> diag(A)<- rep(-1,times=5)
R> b <- rep(0,times=5)
R> res <- CEoptim(dirichletLoglike,f.arg=list(Y=y,n=100,K=4),maximize=T,
       continuous=list(mean=mu0,sd=sigma0,conMat=A,conVec=b,smoothSd=0.5),
       N=10000L,verbose=TRUE)
\end{CodeInput}

\begin{figure}[htb]
\centerline{\includegraphics[width=0.7\linewidth]{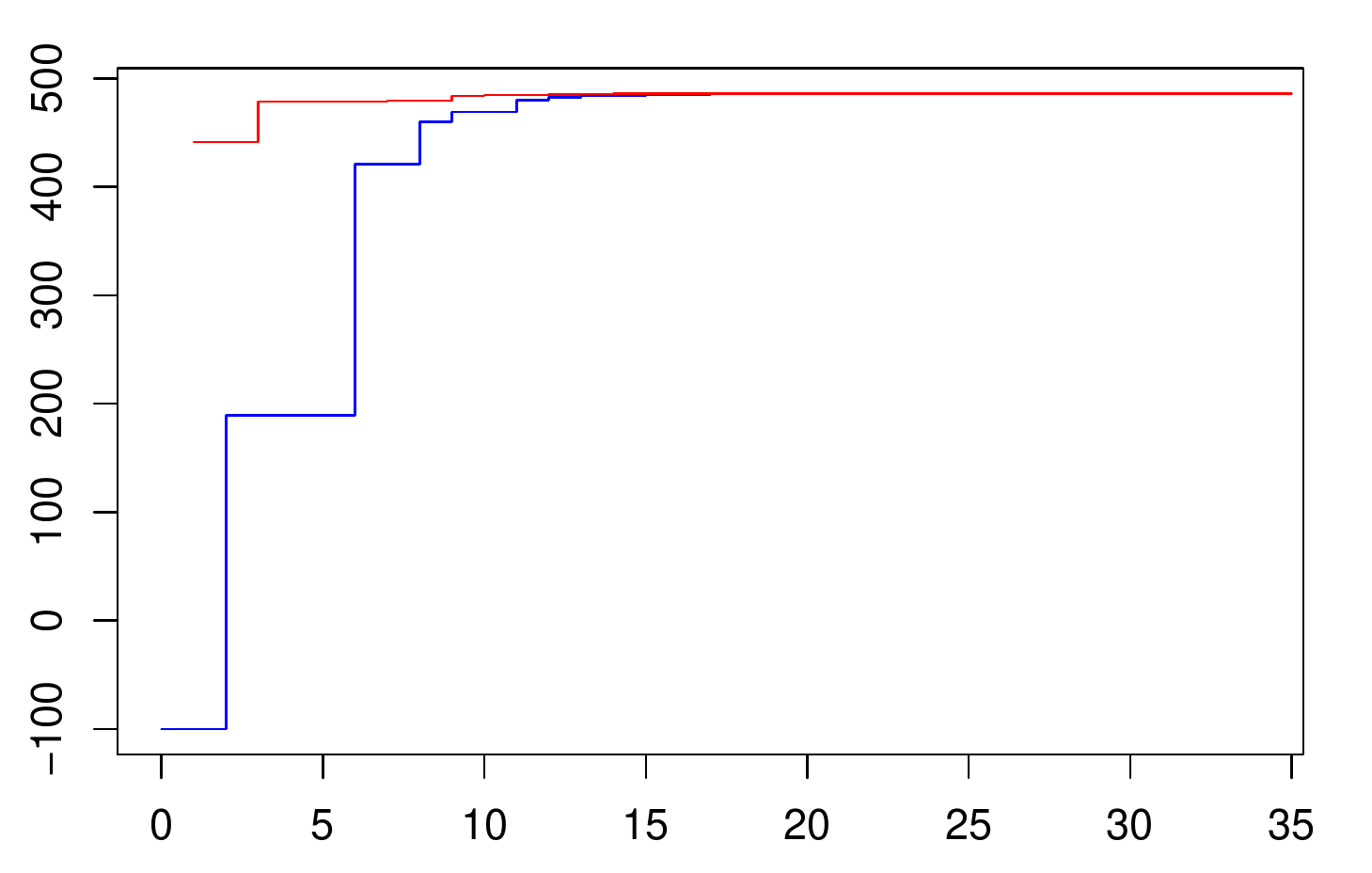}}
\caption{Evolution of the best value (upper line) and the worst value of the
  best (elite) samples (lower line) }
\label{fig:ceevolution}
\end{figure}

With the returned \code{states} variable, we can plot the
evolution of optimal values per iteration, as shown in Figure
\ref{fig:ceevolution}, where the upper line indicates the
best value found so far, while the lower line gives the worst value of
the current elite sample.
\begin{CodeInput}
R> par(mai=c(0.6,1,0.5,0.2),oma=c(0,0,0,1))
R> plot(res$states[,'iter'],res$states[,'gammat'],type='s',col="blue",xlab="",ylab="")
R> lines(res$states[,'optimum'],type='s',col="red")
\end{CodeInput}

\begin{CodeInput}
R> res
\end{CodeInput}
\begin{CodeOutput}
Optimizer for continuous part: 
 1.111656 2.000186 3.534268 3.983616 5.142336 
Optimum: 
 486.2124 
Number of iterations: 
 35 
Convergence: 
 Variance converged
\end{CodeOutput}
Maximum likelihood estimates for Dirichlet data can be computed to
high accuracy
via  the fixed-point techniques of
\cite{estDirichlet}. This requires sophisticated
numerical techniques for inverting
digamma functions.
When applying this  method to the same
$\Dirichlet(1,2,3,4,5)$ data, we obtained the estimate
$\hat{\balpha}=(1.111715, 2.000243,$ $3.534321,$ $3.983752,$
$5.142596)$, with a likelihood value of $486.2124$, giving excellent
agreement between the two approaches.
%\intrusion{Check what the function value is for both of them}

\subsection{Lasso regression} \label{eg:Lasso}
\rm Suppose that we observed some data from the following model:
$$Y_i=\bx_i^\top\boldsymbol{\beta}+\epsilon_i,\ \ \ i=1,\ldots,n\;,$$
where $\bx_i=(x_{i1},\ldots,x_{ip})^\top$ is the $p$-vector
of explanatory variables,
$\boldsymbol{\beta}=(\beta_1,\ldots,\beta_p)^\top$ is the $p$-vector of
regression coefficients, and the $\{\epsilon_i\}$ are the noise terms
with
$\Em[\epsilon_i]=0$, $\Var[\epsilon_i]=\sigma^2$, for all $i$ and
$\Cov(\epsilon_i,\epsilon_j)=0$ ($\forall i \neq j $). Consider a
Lasso regression approach to estimate the regression vector
$\vect{\beta}$:
\begin{eqnarray*}
\hat{\vect{\beta}}^{\text{lasso}}&=& \underset{\vect{\beta}\in \R^p}{\argmin}\frac{1}{2n}\sum_{i=1}^n(Y_i-\bx_i^\top\vect{\beta})^2+\lambda\sum_{j=1}^p|\beta_j|\\
&=& \underset{\vect{\beta}\in
  \R^p}{\argmin}\underbrace{\frac{1}{2n}\|\boldsymbol{Y}-\mathcal{X}\vect{\beta}\|^2_2}_{\text{Loss}}+\lambda\underbrace{\|\vect{\beta}
  \|_1}_{\text{Penalty}},
\end{eqnarray*}
where $\boldsymbol{Y}=(Y_1,\ldots,Y_n)^\top$ and $\mathcal{X}=(\bx_1,\ldots,\bx_n)^\top$ is the $(n\times p)$ design matrix. The tuning parameter $\lambda$ controls the  amount of regularization.
%%\intrusion{Do we want to consider the actual optimization problem?}
%% An equivalent formulation of this optimisation problem is:
%% $$\underset{\vect{\beta}\in  \R^p}{\text{min}}\sum_{i=1}^n\left(Y_i-\sum_{j=1}^p\beta_jX_{ij}\right)^2\ \ \ \ \ \text{subject to } \sum_{j=1}^p|\beta_j|\leq t.$$

For a given value of $\lambda$, we will use CE method to obtain the Lasso regression coefficient and
compared our results with those obtained by the function
\code{glmnet} from the package \pkg{glmnet} presented by  \cite{friedman2008}.

 We 
generate data of size $n=150$, with $p=60$ explanatory variables
independently generated from a standard normal distribution. The true coefficients from
$\vect{\beta}$ are chosen such that 10 are large (between 0.5 and 1)
and 50 are exactly 0. The variance of the noise is equal to 1.
\begin{CodeInput}
R> set.seed(10)
R> n <- 150
R> p <- 60
R> beta <- c(runif(10,0.5,1),rep(0,50))
R> X <- matrix(rnorm(n*p),ncol=60)
R> Y <- X%*%matrix(beta,ncol=1)+rnorm(n) 
\end{CodeInput}
We first use the \code{glmnet} function to find the Lasso regression
coefficient that gives a {\em sparsity} of 10; that is, exactly 10
coefficients are non-zero.
%\intrusion{Say here that you want a $\lambda$ that gives a
%  ``sparsity'' parameter of 10. What does that actually mean?}
%Note that the\code{glmnet} function looking for the minimum of$\frac{\text{RSS}}{2n} + \lambda   ||\vect{\beta}||_1$.
%%\intrusion{So why not the original minimization problem? Or actually,
 %%the lambda's differ by a factor of 2. Maybe emphasize that.}
%%\intrusion{Also probably better to specify what \text{RSS} is, althoug
 %% that should be automatic for statisticians}
\begin{CodeInput}
R> require(glmnet)
R> res.glmnet <- glmnet(X,Y)
# Find the lambda value to get a model with a sparsity=10
R> sparsity.10 <- which(res.glmnet$df==10)
R> (lambda.10 <- res.glmnet$lambda[sparsity.10[1]])
\end{CodeInput}
\begin{CodeOutput}
0.2731371
\end{CodeOutput}
\begin{CodeInput}
R> beta.glmnet <- res.glmnet$beta[,sparsity.10[1]]
\end{CodeInput}
The corresponding indices are correctly identified by \code{glmnet}:
\begin{CodeInput}
# Index of the non-zero coefficient
R> (ind.beta <- which(res.glmnet$beta[,sparsity.10[1]]!=0))
\end{CodeInput}
\begin{CodeOutput}
V1  V2  V3  V4  V5  V6  V7  V8  V9 V10
  1   2   3   4   5   6   7   8   9  10
\end{CodeOutput}
\begin{CodeInput}
# Values of the non-zero coefficient (NZ)
R> (beta.glmnet.NZ <- res.glmnet$beta[ind.beta,sparsity.10[1]])
\end{CodeInput}
\begin{CodeOutput}
       V1         V2         V3         V4         V5         V6
0.39006188 0.39345242 0.40795534 0.57510345 0.18776598 0.19553092
        V7         V8         V9        V10
0.02929225 0.55435619 0.57656731 0.56279719
\end{CodeOutput}

We now use our function to estimate the Lasso regression function for the given $\lambda=0.2731371$.
\begin{CodeInput}
R> require(CEoptim)
R> RSS.penalized <- function(x,X,Y,lambda){
     out <- (1/2)*mean((Y-X%*%matrix(x,ncol=1,nrow=dim(X)[2],byrow=TRUE))**2)
     + lambda*sum(abs(x))
     return(out)}

R> mu0 <- rep(0,times=p)
R> sigma0 <- rep(5,times=p)
R> N <- 1000
R> set.seed(1212)

R> res <- CEoptim(RSS.penalized,f.arg=list(X=X,Y=Y,lambda=lambda.10),
                   continuous=list(mean=mu0,sd=sigma0,sdThr=0.00001),N=N)
R> beta.CEoptim <- res$optimizer$continuous
R> # Index of the non-zero coefficient
R> (ind.beta.CEoptim.NZ <- which(abs(beta.CEoptim)>0.000001))
\end{CodeInput}
\begin{CodeOutput}
[1]  1  2  3  4  5  6  7  8  9 10
\end{CodeOutput}
\begin{CodeInput}
R> beta.CEoptinm.NZ <- beta.CEoptim[ind.beta.CEoptim.NZ]
R> (compare.beta.NZ <- rbind(beta.glmnet.NZ,beta.CEoptinm.NZ))
\end{CodeInput}
\begin{CodeOutput}
                        V1        V2        V3        V4        V5        V6
beta.glmnet.NE   0.3900619 0.3934524 0.4079553 0.5751035 0.1877660 0.1955309
beta.CEoptinm.NE 0.3631798 0.3826273 0.4419025 0.6014707 0.1639559 0.1721753
                          V7        V8        V9       V10
beta.glmnet.NE   0.029292247 0.5543562 0.5765673 0.5627972
beta.CEoptinm.NE 0.005821685 0.5537388 0.5854710 0.6034473
\end{CodeOutput}
The two methods give similar values for the non-zero coefficient,
although they are not
exactly the same. Note, however, that of the two solutions the one
found by
 \code{CEoptim} gives the smaller value for the objective
function $\frac{\text{RSS}}{2n} + \lambda   ||\vect{\beta}||_1$ (where Residual Sum of Square: RSS$=\|\boldsymbol{Y}-\mathcal{X}\vect{\beta}\|^2_2$).

\begin{CodeInput}
R> (RSS.penalized(beta.CEoptim,X=X,Y=Y,lambda=lambda.10))
\end{CodeInput}
\begin{CodeOutput}
[1] 1.990268
\end{CodeOutput}

\begin{CodeInput}
R> (RSS.penalized(beta.glmnet,X=X,Y=Y,lambda=lambda.10))
\end{CodeInput}
\begin{CodeOutput}
[1] 1.993622
\end{CodeOutput}

\begin{figure}[H]
\label{fig:lasso}
%\centerline{\includegraphics[width=0.8\linewidth]{figure-lasso.png}}
\centerline{\includegraphics[width=0.95\linewidth]{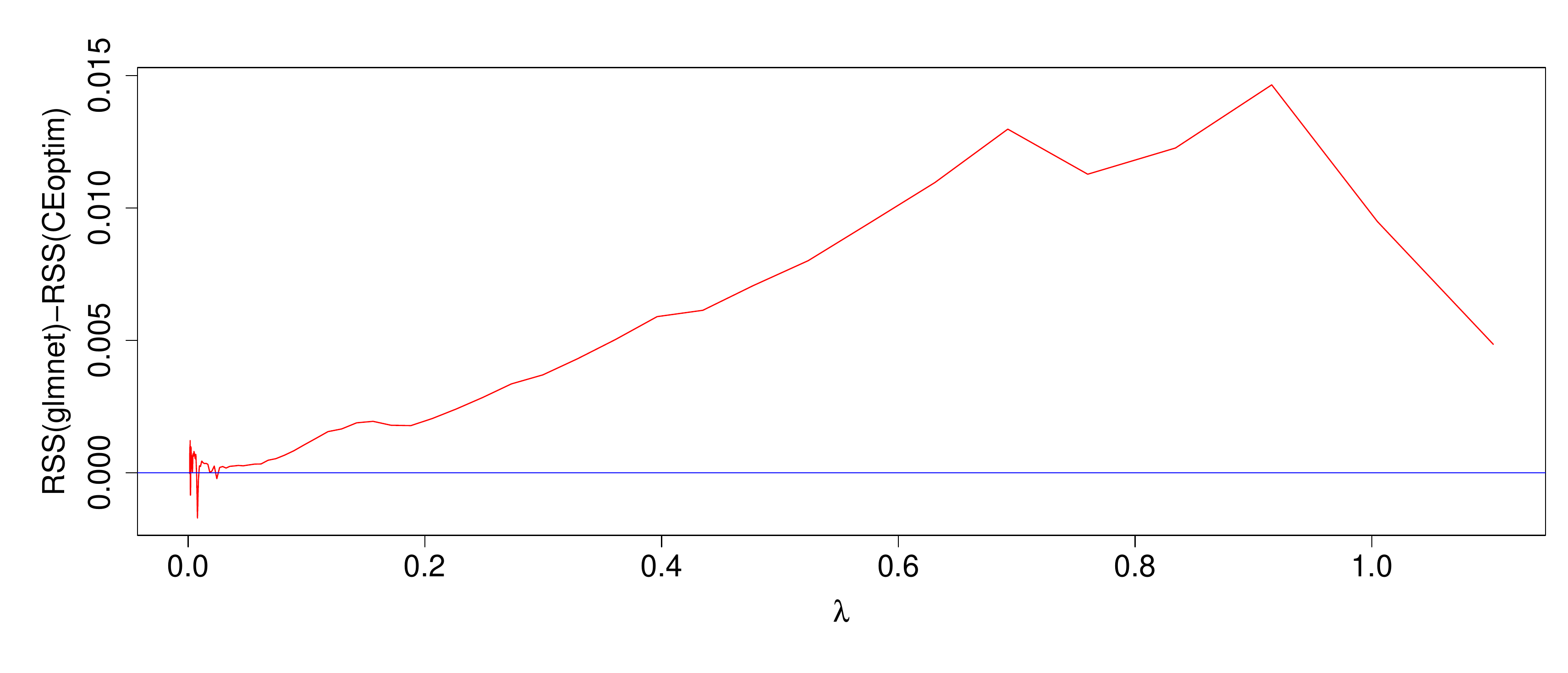}}
\caption{Difference of the objective function values between \code{glmnet} and  \code{CEoptim}  for a sequence of 73 values of $\lambda$.
}
\end{figure}

Further, we compare the results obtained by \code{CEoptim} with the
ones given by \code{glmnet} for the sequence of tuning parameter
$\lambda$ used by default in  the \code{glmnet} function. Results given by  \code{CEoptim}  are slightly better than \code{glmnet}  optimizer (see Figure \ref{fig:lasso}).
In more than 90\% of the cases (over the 73 values of $\lambda$
investigated) \code{CEoptim} gives a lower value for the objective function than \code{glmnet}. However, the coordinate descent algorithm \citep{friedman2008} used in \code{glmnet} is computationally less demanding than the CE approach.

%\intrusion{Make the axes font bigger. Also, it is difficult
%  compare. Maybe plot the difference too!}

\subsection{AR(1) model with regime switching}

As a final  illustration of the use of \pkg{CEoptim}, we consider a
model fitting problem involving
both continuous and discrete variables.

Let $Y_t$ be the added value of a stock at time $t$, at day
$t=1,2,\ldots,300$; that is, the increase (which may be negative) in
stock price relative to the price at
time $t=0$.  Let $X_t$ be the increment at day $t$. Hence,
\[
Y_t = \sum_{i=1}^{t} X_i, \quad t=1,\ldots,300.
\]
We assume that the $\{X_i\}$ satisfy a zero-mean AR(1) model with
three possibly different regimes. Specifically, we assume
\[
X_i =
\theta_i \, X_{i-1} + \epsilon_i, \quad i=1,\ldots,300\;,
\]
where
\begin{equation}\label{thetar}
\theta_i =
\begin{cases}
\theta^{(1)}\;, \quad i=1,\ldots,r_1 \\
\theta^{(2)}\;,  \quad i=r_1+1,\ldots,r_2 \\
\theta^{(3)} \;, \quad i=r_2+1,\ldots,300\;,
\end{cases}
\end{equation}

$1 \leq r_1 < r_2 < 300$, $|\theta_i| \leq 1$,
$i=1,2,3$,  and the error terms $\{\epsilon_i\}$ are iid and normally
distributed with standard deviation $\sigma$. The model thus has two
discrete and three continuous parameters, as well as a nuisance
parameter $\sigma$. Define
$\vect{\theta} = (\theta^{(1)},\ \theta^{(2)},\theta^{(3)})^\top$,
$\br = (r_1,r_2)^\top$, and
let $x_1,\ldots,x_{300}$ be the
observed increments. We put  $x_0 = 0$.
We fit the parameters by minimizing the least squares function
\[
L(\vect{\theta},\br) = \sum_{i=1}^{300} (x_i - \hat{x}_i)^2\;,
\]
where $\hat{x}_i$ is the fitted value $\theta_i \, x_{i-1}$, and
$\theta_i$ is determined by $\vect{\theta}$ and $\br$ via
\eqref{thetar}. The vector of fitted values, say $\hat{\bx}$, can be
written in matrix notation as $\hat{\bx} = \mathcal{X} \vect{\theta}$, where
 $\mathcal{X}$ is a $300 \times 3$ matrix where
the elements in rows $1,\ldots,r_1$ in the first column are equal to
$x_0,\ldots,x_{r_1-1}$; the elements in rows $r_1+1,\ldots,r_2$ in the
second column are equal to $x_{r_1},\ldots,x_{r_2 - 1}$; the elements in
rows $r_2+1,\ldots,300$ in the third column are equal to
$x_{r_2},\ldots,x_{299}$; and all other elements are 0. 
The implementation of the least squares 
function is given below. Note that the
function requires input $\br -1$ rather than $\br$, because
each categorical variable used in \code{CEoptim} takes value in a set
$\{0,\ldots,c\}$ for some $c$.

\begin{CodeInput}
R> sumsqrs <- function(theta,rm1,x) {
      N <- length(x) #without x[0]
      r <- 1 + sort(rm1)  # internal end points of regimes
      if (r[1]==r[2]) { # test for invalid regime
        return(Inf);
      }
   thetas <- rep(theta, times=c(r,N)-c(1,r+1)+1)
   xhat <- c(0,head(x,-1))*thetas
   # Compute sum of squared errors
   sum((x-xhat)^2)
   }
\end{CodeInput}

The data have been generated using the parameters
$\vect{\theta} = (0.3,0.9,-0.9)$, $\br = (100,200)$ and $\sigma= 0.1$.
The data are included in the package and are available by using: 
\begin{CodeInput}
R> data(yt)
R> xt <- yt - c(0,yt[-300])
\end{CodeInput}
The following code implements the use of \pkg{CEoptim} for this
constrained mixed problem. 
\begin{CodeInput}
R> A <- rbind(diag(3),-diag(3))
R> b <- rep(1,6)
R> set.seed(123)
R> require(CEoptim)
R> res <- CEoptim(f=sumsqrs, f.arg=list(xt), continuous=list(mean=c(0,0,0),
         sd=rep(1.0,3), conMat=A, conVec=b),discrete=list(categories=c(298L,298L),
         smoothProb=0.5),N=10000,rho=0.001, verbose=TRUE)
\end{CodeInput}

\begin{figure}[H]
\centerline{\includegraphics[width=1\linewidth]{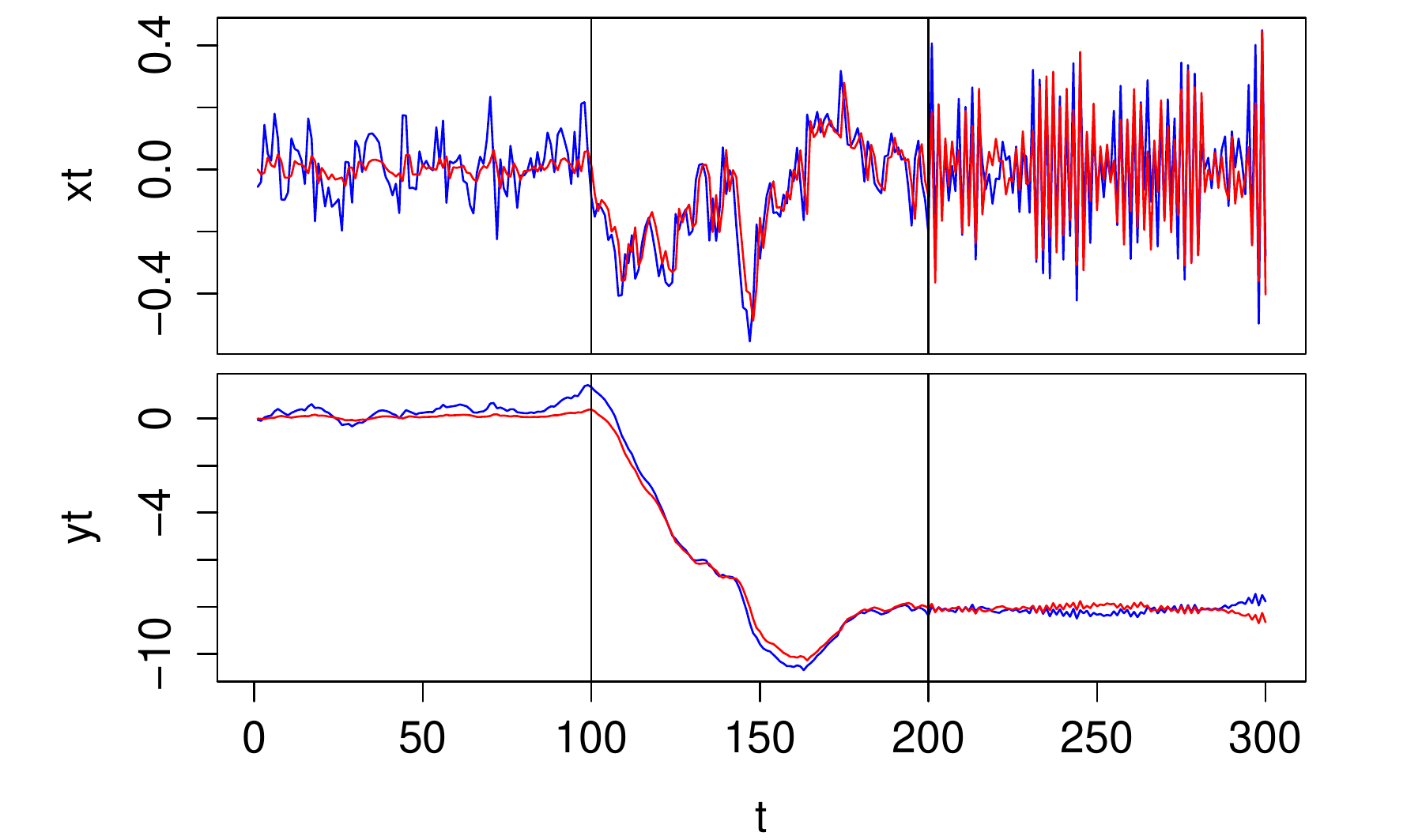}}
\caption{Assessment of the fit of the model: $x_t$ (top) and $y_t$ (bottom).}
\label{fig:regimedif}
\end{figure}

The output is as follows:
\begin{CodeInput}
R> res
\end{CodeInput}
\begin{CodeOutput}
Optimizer for continuous part: 
 0.2702714 0.8801672 -0.8975874 
Optimizer for discrete part: 
 99 199 
Optimum: 
 2.675727 
Number of iterations: 
 13 
Convergence: 
 Variance converged 
\end{CodeOutput}
As the input to \code{CEoptim} is $\br -1$,  the optimal
vector $\br$ is given by 
\begin{CodeInput}
R> (est.r <- sort(res$optimizer$discrete)+1)
\end{CodeInput}
\begin{CodeOutput}
[1] 100 200
\end{CodeOutput}
which gives exactly the ``true'' boundaries for the regimes.
From the estimates of the model, one can assess the fit of the model by comparing $y_t$ with $\hat{y}_t=\sum_{i=1}^{300}\hat{x_t}$
and $x_t$ against the fit
$\hat{x}_t$. Figure~\ref{fig:regimedif}  shows an excellent fit.

\begin{CodeInput}
R> t <- 1:300
R> est.theta <- res$optimizer$continuous
R> est.thetas <- rep(est.theta,times=c(est.r,300) - c(1,est.r+1) + 1)
R> xfit <- c(0,head(xt,-1))*est.thetas

R> par(mfrow=c(2,1))
R> plot(xt~t,type="l",col="blue")
R> lines(xfit,col="red")
R> abline(v=c(100,200))
R> plot(yt,type="l",col="blue")
R> lines(cumsum(xfit),col="red")
R> abline(v=c(100,200))
\end{CodeInput}

\begin{figure}[H]
\centerline{\includegraphics[width=1\linewidth]{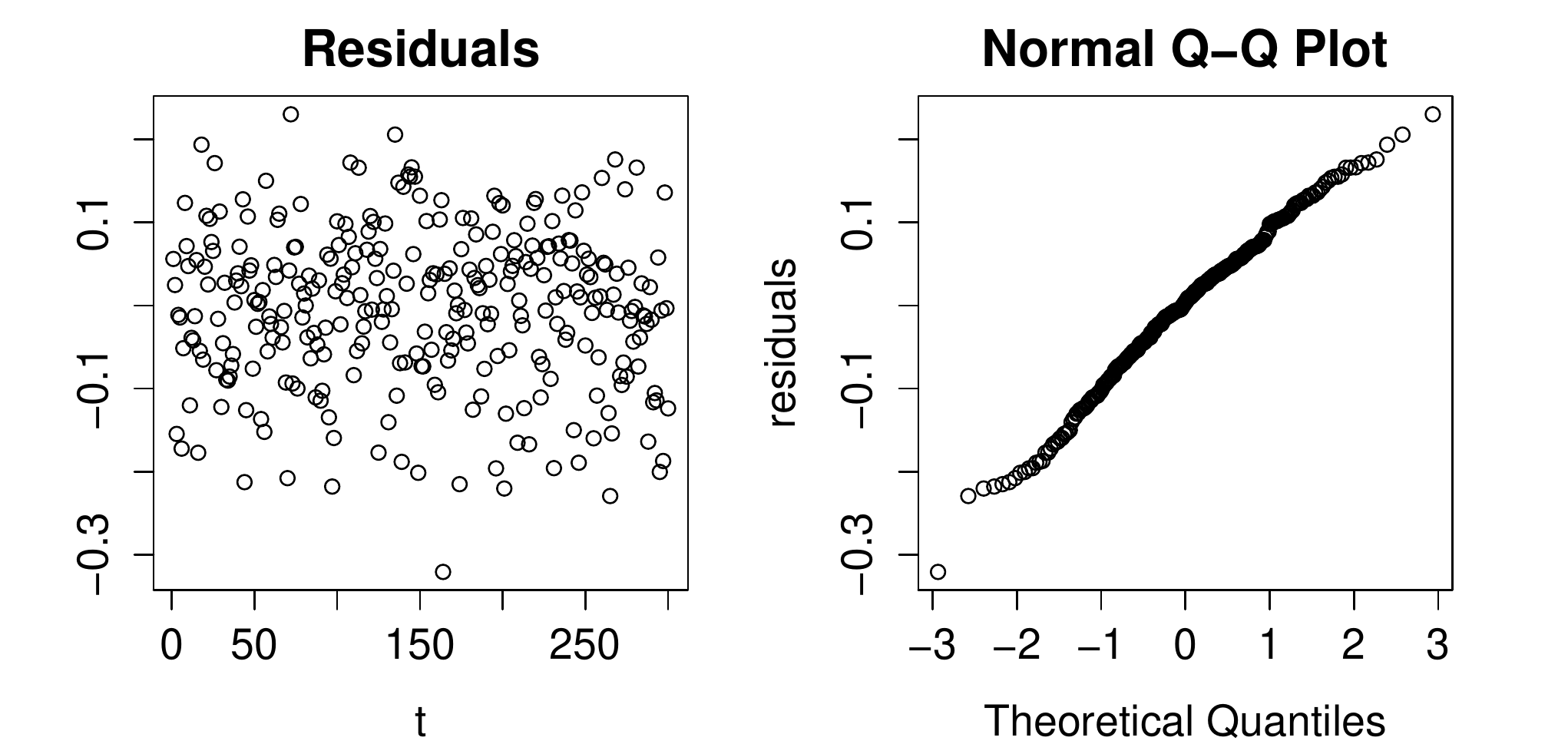}}
\caption{Diagnostic residuals of the model: scatterplot of the
  residuals  (left) and  quantile quantile normal plot(right).}
\label{fig:residuals}
\end{figure}

A diagnostic of the residuals is presented in
Figures~\ref{fig:residuals}, showing a normal quantile plot (left panel) and a scatterplot of the residuals (right panel). 
\begin{CodeInput}
R> par(mfrow=c(1,2))
R> resid <- xfit - xt
R> plot(resid,ylab="residuals",xlab="t")
R> qqnorm(resid,ylab="residuals")
\end{CodeInput}

%\pkg{CEoptim} optimiser fits well the model  according to Figures~\ref{fig:regimedif} and Figures~\ref{fig:residuals}.

\section{Concluding remarks}

\pkg{CEoptim} provides the \proglang{R} implementation of the
cross-entropy method for optimization.  The  versatility and
effectiveness of this new package have been
illustrated through a variety of optimization example, involving continuous,
discrete, mixed and constrained optimization problems.
We have  demonstrated how this simple algorithm can be of benefit in
statistical inference, including model fitting, regression,
maximum likelihood, and lasso methods. \pkg{CEoptim} is available from
the Comprehensive \proglang{R} Archive Network (CRAN) at
\url{http://cran.r-project.org/}.

\section*{Acknowledgments}

This work was supported by the Australian Research Council {\em Centre
  of Excellence for Mathematical and Statistical Frontiers} (ACEMS)
under grant number CE140100049. Qibin Duan would also like to
acknowledge the support from the University of Queensland through the
UQ International Scholarships scheme.

%\intrusion{We don't need 2 Sani references. Only 2008 reference.}
\bibliography{biblio}
\end{document}

%% file: definitions.tex
\newtheorem{remark}{Remark}

\newcommand{\ealg}{\end{alg}}
\newcommand{\balg}{\begin{alg}}

\newcommand{\ceil}[1]{\left\lceil #1 \right\rceil}

\newcommand{\KL}{{\EuScript{D}}}

\newcommand{\bmu}{\vect{\mu}}

\newcommand{\bX}{\mathbf{X}}
\newcommand{\bx}{\mathbf{x}}

\newcommand{\br}{\mathbf{r}}
\newcommand{\by}{\mathbf{y}}

\newcommand{\bu}{\mathbf{u}}

\newcommand{\bY}{\mathbf{Y}}

\renewcommand{\tilde}{\widetilde}

\newcommand{\balpha}{\vect{\alpha}}

\newcommand{\bfb}{\mathbf{b}}

\newcommand{\Var}{\mathrm{Var}}

\newcommand{\Cov}{\mathrm{Cov}}

\newcommand{\matlab}{\mathrm{M}\mathrm{{\scriptstyle ATLAB}}}

\newcommand{\bp}{\mathbf{p}}

\newcommand{\bv}{\mathbf{v}}

\renewcommand{\epsilon}{\varepsilon}
\renewcommand{\rho}{\varrho}
\renewcommand{\log}{\ln}
\renewcommand{\hat}{\widehat}
\renewcommand{\leq}{\leqslant}
\renewcommand{\geq}{\geqslant}

\newcommand{\argmin}{\mathop{\rm argmin}}

\newcommand{\elite}{\mathop{\rm e}}

\newcommand{\Ber}{{\sf Ber}}
\newcommand{\ber}{\Ber}

\newcommand{\Nor}{{\sf N}}

\newcommand{\I}{\mathrm{I}}

\newcommand{\Dirichlet}{{\sf Dirichlet}}

\newcommand{\Em}{\mathbb E}
\newcommand{\Pm}{\mathbb P}
\newcommand{\R}{\mathbb R}

\newcommand{\scE}{\mathscr{E}}

\newcommand{\scX}{\mathscr{X}}

\newcommand{\e}{\mathrm{e}}

\newcommand{\vect}[1]{\boldsymbol #1}

\newcommand{\di}{\mathrm{d}}

\def\acro#1#2{\vskip4pt\hbox to\textwidth{\normalsize
\hbox to5pc{#1\hfill}\vtop{\advance\hsize by
-5pc\raggedright\noindent#2}}}

\def\symbol#1#2{\vskip4pt\hbox to\textwidth{\normalsize
\hbox to5pc{#1\hfill}\vtop{\advance\hsize by
-5pc\raggedright\noindent#2}}}

\def\distro#1#2{\vskip4pt\hbox to\textwidth{\normalsize
\hbox to5pc{#1\hfill}\vtop{\advance\hsize by
-5pc\raggedright\noindent#2}}}

\newcommand{\simiidt}{{\:{\sim}_{\mathrm{iid}}\:}}

\newcommand{\chk}[1]{}

%% file: CEoptim_arxiv.bbl
\begin{thebibliography}{20}
\newcommand{\enquote}[1]{``#1''}
\providecommand{\natexlab}[1]{#1}
\providecommand{\url}[1]{\texttt{#1}}
\providecommand{\urlprefix}{URL }
\expandafter\ifx\csname urlstyle\endcsname\relax
  \providecommand{\doi}[1]{doi:\discretionary{}{}{}#1}\else
  \providecommand{\doi}{doi:\discretionary{}{}{}\begingroup
  \urlstyle{rm}\Url}\fi
\providecommand{\eprint}[2][]{\url{#2}}

\bibitem[{Alon \emph{et~al.}(2005)Alon, Kroese, Raviv, and
  Rubinstein}]{alon2005allocation}
Alon G, Kroese DP, Raviv T, Rubinstein RY (2005).
\newblock \enquote{Application of the cross-entropy method to the buffer
  allocation problem in a simulation-based environment.}
\newblock \emph{Annals of Operations Research}, \textbf{134}(1), 137--151.

\bibitem[{Bendtsen(2012)}]{bendtsenpso}
Bendtsen C (2012).
\newblock \emph{pso: Particle Swarm Optimization}.
\newblock R package version 1.0. 3,
  \urlprefix\url{http://CRAN.R-project.org/package=pso}.

\bibitem[{Botev \emph{et~al.}(2013)Botev, Kroese, Rubinstein, and
  L'Ecuyer}]{ceoptbotev2013}
Botev ZI, Kroese DP, Rubinstein RY, L'Ecuyer P (2013).
\newblock \enquote{The cross-entropy method for optimization.}
\newblock \emph{Machine Learning: Theory and Applications, V. Govindaraju and
  C.R. Rao, Eds, Chennai: Elsevier B.V.}, \textbf{31}, 35--59.

\bibitem[{De~Boer \emph{et~al.}(2005)De~Boer, Kroese, Mannor, and
  Rubinstein}]{cetutorial2005}
De~Boer PT, Kroese DP, Mannor S, Rubinstein RY (2005).
\newblock \enquote{A tutorial on the cross-entropy method.}
\newblock \emph{Annals of operations research}, \textbf{134}(1), 19--67.

\bibitem[{Duan \emph{et~al.}(2014)Duan, Kroese, Brereton, Spettl, and
  Schmidt}]{duan2014inverting}
Duan Q, Kroese DP, Brereton T, Spettl A, Schmidt V (2014).
\newblock \enquote{Inverting Laguerre Tessellations.}
\newblock \emph{The Computer Journal}, \textbf{57}, 1431--1440.

\bibitem[{Friedman \emph{et~al.}(2008)Friedman, Hastie, and
  Tibshirani}]{friedman2008}
Friedman J, Hastie T, Tibshirani R (2008).
\newblock \enquote{Regularization Paths for Generalized Linear Models via
  Coordinate Descent.}
\newblock \emph{Journal of Statistical Software}, \textbf{33}.

\bibitem[{Knuth(1993)}]{lesmis_knuth}
Knuth DE (1993).
\newblock \emph{The Stanford GraphBase: A Platform for Combinatorial
  Computing}.
\newblock ACM Press, Reading,MA.

\bibitem[{Kobilarov(2012)}]{kobilarov2012motionplanning}
Kobilarov M (2012).
\newblock \enquote{Cross-entropy motion planning.}
\newblock \emph{The International Journal of Robotics Research},
  \textbf{31}(7), 855--871.

\bibitem[{Kothari and Kroese(2009)}]{kothari09integerprograming}
Kothari RP, Kroese DP (2009).
\newblock \enquote{Optimal generation expansion planning via the cross-entropy
  method.}
\newblock In \emph{Winter Simulation Conference}, pp. 1482--1491.

\bibitem[{Kroese \emph{et~al.}(2007)Kroese, Hui, and
  Nariai}]{kroese2007network}
Kroese DP, Hui KP, Nariai S (2007).
\newblock \enquote{Network reliability optimization via the cross-entropy
  method.}
\newblock \emph{Reliability, IEEE Transactions on}, \textbf{56}(2), 275--287.

\bibitem[{Minka(2000)}]{estDirichlet}
Minka TP (2000).
\newblock \enquote{Estimating a Dirichlet distribution.}
\newblock \emph{Technical report}, M.I.T.
\newblock
  \urlprefix\url{http://research.microsoft.com/en-us/um/people/minka/papers/dirichlet/}.

\bibitem[{Mullen \emph{et~al.}(2011)Mullen, Ardia, Gil, Windover, and
  Cline}]{mullendeoptimJss}
Mullen KM, Ardia D, Gil DL, Windover D, Cline J (2011).
\newblock \enquote{DEoptim: An R package for global optimization by
  differential evolution.}
\newblock \emph{Journal of Statistical Software}, \textbf{40}.

\bibitem[{Nagumo \emph{et~al.}(1962)Nagumo, Arimoto, and Yoshizawa}]{nagumo62}
Nagumo J, Arimoto S, Yoshizawa S (1962).
\newblock \enquote{An active pulse transmission line simulating nerve axon.}
\newblock \emph{Proceedings of the IRE}, \textbf{50}(10), 2061--2070.

\bibitem[{Ramsay \emph{et~al.}(2007)Ramsay, Hooker, Campbell, and
  Cao}]{Ramsay07}
Ramsay JO, Hooker G, Campbell D, Cao J (2007).
\newblock \enquote{Parameter estimation for differential equations: A
  generalized smoothing approach.}
\newblock \emph{Journal of the Royal Statistical Society, Series B},
  \textbf{69}(5), 741--796.

\bibitem[{Rubinstein(1997)}]{rubinstein1997}
Rubinstein RY (1997).
\newblock \enquote{Optimization of computer simulation models with rare
  events.}
\newblock \emph{European Journal of Operational Research}, \textbf{99}(1),
  89--112.

\bibitem[{Rubinstein(1999)}]{rubinstein1999}
Rubinstein RY (1999).
\newblock \enquote{The cross-entropy method for combinatorial and continuous
  optimization.}
\newblock \emph{Methodology and computing in applied probability},
  \textbf{1}(2), 127--190.

\bibitem[{Rubinstein and Kroese(2004)}]{cebook}
Rubinstein RY, Kroese DP (2004).
\newblock \emph{The Cross-Entropy Method: A Unified Approach to Combinatorial
  Optimization, Monte-Carlo Simulation and Machine Learning}.
\newblock Springer, New York.

\bibitem[{Rubinstein and Kroese(2008)}]{mcbook}
Rubinstein RY, Kroese DP (2008).
\newblock \emph{Simulation and the Monte Carlo method}.
\newblock 2nd edition. John Wiley \& Sons, New York.

\bibitem[{Sani and Kroese(2008)}]{sani2008controlling}
Sani A, Kroese DP (2008).
\newblock \enquote{Controlling the number of HIV infectives in a mobile
  population.}
\newblock \emph{Mathematical biosciences}, \textbf{213}(2), 103--112.

\bibitem[{Xiang \emph{et~al.}(2013)Xiang, Gubian, Suomela, and
  Hoeng}]{xiang2013sa}
Xiang Y, Gubian S, Suomela B, Hoeng J (2013).
\newblock \enquote{Generalized simulated annealing for global optimization: the
  GenSA Package.}
\newblock \emph{R Journal}, \textbf{5}(1).

\end{thebibliography}
